\documentclass[aps,pra,twocolumn,showpacs,preprintnumbers,amsmath,amssymb]{revtex4-1}
\usepackage{graphicx}
\usepackage[usenames,dvipsnames]{xcolor}
\usepackage{hyperref}
\usepackage{cleveref}
\usepackage{makecell}
\usepackage{float}
\usepackage{lineno}
\usepackage[T1]{fontenc}
\usepackage[caption=false]{subfig}
\newcommand{\Tr}{\textrm{Tr}}

\newcommand{\bra}[1]{\ensuremath{\langle #1 |}}
\newcommand{\ket}[1]{\ensuremath{| #1 \rangle}}

\usepackage{amssymb,amsmath} 
\usepackage{amsfonts}
\usepackage{color}
\usepackage[utf8]{inputenc}
\usepackage{lmodern}
\usepackage{epsfig}
\usepackage{ulem}
\usepackage{blindtext}
\begin{document}
\title{Ground-state phase diagram of Rydberg atoms in a triangular-prism array}

\author{Qing-Yuan Zuo$^1$}
\author{Shuo Geng$^2$}
\author{Shan-Wen Tsai$^2$}
\author{Jin Zhang$^1$}
\email{jzhang91@cqu.edu.cn}
\affiliation{$^1$Department of Physics and Chongqing Key Laboratory for Strongly Coupled Physics, Chongqing University, Chongqing 401331, People's Republic of China}
\affiliation{$^2$ Department of Physics and Astronomy, University of California, Riverside, California 92521, USA}

\definecolor{burnt}{cmyk}{0.2,0.8,1,0}
\def\lt{\lambda ^t}
\def\note{note}
\def\beq{\begin{equation}}
\def\enq{\end{equation}}

\date{\today}
\begin{abstract}
We study the ground-state phase diagram of Rydberg atoms in a triangular-prism optical tweezer array using the density matrix renormalization group. By tuning the detuning–Rabi-frequency ratio and the Rydberg blockade radius, the system realizes multiple density-wave phases characterized by spontaneous breaking of translational and leg-exchange symmetries. Unlike two-leg Rydberg ladders with $\mathbb{Z}_2$ leg-exchange symmetry, the triangular prism possesses $\mathbb{D}_3$ symmetry, resulting in a richer set of ordered phases and transitions. For blockade radius moderately larger than the lattice spacing, a density-wave phase with alternating double and single Rydberg occupancy appears at large detuning. It breaks $\mathbb{Z}_2$ translational and $\mathbb{Z}_3$ rotational symmetry while preserving the reflection symmetry of the triangular rung. Upon decreasing detuning, it melts via a Berezinskii–Kosterlitz–Thouless (BKT) transition into a critical phase and subsequently enters a disordered phase through a second BKT transition, forming a commensurate regime described at low energies by a $\mathbb{Z}_6$ clock model. At larger blockade radius, a phase with one Rydberg excitation per triangle that breaks $\mathbb{D}_3$ symmetry emerges through a first-order transition. When double excitations on neighboring triangles are suppressed, rung-trimerized density waves develop as detuning increases from the disordered phase. These phases break translational symmetry while preserving $\mathbb{D}_3$ symmetry, and their melting transitions follow the same critical structure as in Rydberg chains and two-leg ladders: $\mathbb{Z}_2$ symmetry breaking yields Ising critical lines, whereas $\mathbb{Z}_3$ and $\mathbb{Z}_4$ symmetry breaking produce chiral critical lines, with three-state Potts and Ashkin–Teller points occurring only along the corresponding commensurate lines. Within the $\mathbb{Z}_2$ rung-trimerized phase, an entanglement-entropy peak signals a crossover regime with enhanced period-2 density modulation, preceding a first-order transition into a $\mathbb{Z}_2 \times \mathbb{D}_3$ symmetry-breaking phase. Floating phases with incommensurate quasi-long-range order appear between trimerized states of different periods.

\end{abstract}

%\pacs{02.30.Ik,05.30.-d,03.75.Kk,05.30.Jp}

\maketitle

%%%%%%%%%%%%%%%%%%%%%%%%%%%%%%%%%%%%%%%%%%%%%%%%%%%%%%%%%%%%%%%%%%%%%%%%%%
\section{Introduction}\label{sec:introduction}
%%%%%%%%%%%%%%%%%%%%%%%%%%%%%%%%%%%%%%%%%%%%%%%%%%%%%%%%%%%%%%%%%%%%%%%%%%

The study of strongly correlated quantum many-body systems continues to be driven by the search for novel phases of matter and unconventional quantum critical phenomena. Quantum simulation provides a controlled route to access such regimes, particularly in settings where classical numerical approaches become prohibitively expensive. Among existing platforms, Rydberg atom arrays have emerged as a powerful and flexible quantum simulator, combining strong and long-range interactions with a high degree of geometric control enabled by optical tweezers \cite{Labuhn2016RydIsing,Bernien2017Dynamics,Keesling2019Kibble,Leseleuc2019topo,Semeghini2021SL,Ebadi2021_256,Pascal2021AF,ChenContinuous2023}. A defining feature of Rydberg systems is the blockade mechanism, which strongly suppresses multiple excitations within a characteristic blockade radius \cite{DudinRabi2012,RydPhysics2018,BrowaeysIndividual2020}. This constraint, together with programmable geometry, naturally leads to strong correlations, crystalline ordering, and nontrivial quantum melting phenomena.

Building on these features, extensive theoretical and experimental efforts have explored quantum criticalities associated with spontaneous symmetry breaking in Rydberg arrays. Ising-type phase transitions have been experimentally observed in one-dimensional (1D) chains and two-dimensional (2D) square lattices \cite{Labuhn2016RydIsing,Bernien2017Dynamics,Ebadi2021_256,Pascal2021AF}. Beyond Ising universality, numerical and analytical studies have predicted Potts, Ashkin–Teller, chiral, Berezinskii–Kosterlitz–Thouless (BKT), and Pokrovsky–Talapov (PT) transitions out of gapless incommensurate floating phases in $1+1$ dimensions \cite{FendleyDWHard2004prb,ChepigaFloating2019prl,ChepigaLifshitz2021prr,Chepiga&Mila2021Kibble,MaceiraChiral2022prr,OBDKTTriRyd2023pra,Chepiga2Component2024prl}, as well as Potts and O($\mathcal{N}$) universality classes in $2+1$ dimensions \cite{RhineCDWSquare2020prl,RhineKagome2021,CXLiTriangular2022,YZGlassy2023prl}. Studies on nontrivial geometries have also revealed deconfined quantum criticality \cite{DQCTriangle2025prl}, supersolid phases \cite{LukasSuperSolid2025pra}, and geometry-dependent phase diagrams in ladder systems \cite{FromholzPhaseDiagTriLadder2022prb,PhaseRyddressedSquareLadder2022prb,eck2023critical,JinPRDCritical2024,ZhangFloating2025NC,ChepigaLadder2025prr,LiaoPhaseRydLadder2025prb}. Beyond conventional density-wave order, frustrated and nonbipartite geometries have been shown to host quantum spin liquids and intrinsic topological order, including $\mathbb{Z}_2$ topological phases and spin-ice–type states realized in blockade-based dimer and trimer constructions and related lattices \cite{Semeghini2021SL,RubenToricCode2021prx,SamajdarZ2Gauge2023prl,RubenUnifySpinLiquids2022prx,KevinSpinLiquids2022prb,TrimerLiquidQuEra2023,SpinIce3DRyd2025prx}. In addition, correlated regimes such as disorder-free glassy behavior \cite{YZGlassy2023prl} and the quantum slush state \cite{ZhangSlush2024prl} have been identified. Rydberg arrays have also become a versatile setting for lattice gauge theories, with proposals and realizations of emergent $\mathbb{Z}_2$, U(1), and $\mathbb{Z}_3$ gauge structures, as well as demonstrations of confinement and string breaking \cite{SuraceLGTString2020prx,Emerge2DGauge2020prx,RealTimeLQED2020prr,FloquetZ2Gauge2024prr,ChengEmergeU1Gauge2024nrp,ObservationStringBreaking2025N,ZhouQSU1Gauge2025cpl,XFZString2026prb}. Recent theoretical and experimental studies have further explored dynamical and nonequilibrium phenomena in programmable Rydberg arrays, including Kibble–Zurek scaling of Ising and Potts criticality and tricritical points \cite{Keesling2019Kibble,NearCriticalKZ2025prl,TriKZLadder2025NC}, quantum many-body scars and revivals in blockade-constrained systems \cite{Bluvstein2021Scars,ScarRydLadder2025prb,ScramblingRyd2025prl,ProxIntegraRevival2025arXiv}, and real-time dynamics of gauge theories \cite{RealTimeLQED2020prr,SuraceLGTString2020prx}. Complementary work has further examined lattice defects and doping in Rydberg arrays \cite{WangDefectsRyd2025prb,DopeAFMRyd2025N,DopedRVBSpinIceRydberg2025pra}.

While the phase diagram of the one-dimensional Rydberg chain is now relatively well understood \cite{MaceiraChiral2022prr,Chepiga2Component2024prl}, and substantial effort has been devoted to understanding quantum phases in two-dimensional and higher-dimensional Rydberg arrays \cite{RhineCDWSquare2020prl,RhineKagome2021,CXLiTriangular2022,DWHoneycomb2022pre,YZGlassy2023prl,TrimerLiquidQuEra2023,OBDKTTriRyd2023pra,ZhangSlush2024prl,DQCTriangle2025prl,XFZString2026prb}, quasi-one-dimensional geometries with multiple coupled chains remain less systematically explored. Rydberg ladders provide the simplest extension beyond a single chain, where blockade constraints, inter-leg couplings, and discrete leg-exchange symmetries interplay to produce a rich phase structure. Recent work in two-leg ladder geometries has uncovered order-by-disorder–induced Ising transitions \cite{SarkarOrderbyDis2023}, tricritical Kibble–Zurek scaling \cite{TriKZLadder2025NC}, revivals and scar-like dynamics \cite{ScarRydLadder2025prb,ProxIntegraRevival2025arXiv}, mappings to effective lattice gauge theories \cite{Emerge2DGauge2020prx,FloquetZ2Gauge2024prr}, and connections to integrable constrained models \cite{EckXXZIntegrableRydLadder2024scipost}. In addition, full phase diagrams featuring density-wave orderings associated with spontaneous breaking of translational and leg-exchange symmetries, rung-dimerized floating phases, and chiral melting transitions out of rung-dimerized density waves have been identified in two-leg ladder systems, with their stability shown to depend sensitively on geometry and aspect ratio \cite{Z4ZigzagLadder2024SciPost,eck2023critical,JinPRDCritical2024,ZhangFloating2025NC,ChepigaLadder2025prr,LiaoPhaseRydLadder2025prb}.

A natural extension beyond ladder geometries is the triangular-prism array, in which three chains are coupled in a way that respects the symmetry of an equilateral triangle. In contrast to two-leg ladders, whose leg-exchange symmetry is $\mathbb{Z}_2$, the triangular prism possesses a $\mathbb{D}_3$ leg-exchange symmetry. This higher discrete symmetry enlarges the space of possible symmetry-breaking patterns and allows for ordered phases and melting transitions without direct analogues in ladder systems. Very recently, prism geometries have begun to be explored in specific contexts, including real-time simulations of $\mathbb{Z}_3$ lattice gauge theories \cite{RealTimeLQED2020prr} and the realization of Potts criticality and tricritical Kibble–Zurek scaling under staggered detuning or staggered atomic species protocols \cite{SarkarOrderbyDis2023,TriKZLadder2025NC}. However, these works focus primarily on dynamical protocols or special parameter regimes. A systematic understanding of the ground-state phase diagram of the Rydberg prism, particularly the competition between translational symmetry breaking and $\mathbb{D}_3$ leg-exchange symmetry breaking across interaction strengths, remains lacking.

In this paper, we study the ground-state phase diagram of Rydberg atoms arranged in a triangular-prism optical tweezer array using the density matrix renormalization group algorithm \cite{DMRG1992prl,DMRG1993prb}. We show that when the Rydberg blockade radius is large enough to suppress double excitations on neighboring triangles, the system forms rung-trimerized density waves as the detuning–Rabi-frequency ratio is increased, in contrast to the rung-dimerized density waves in two-leg ladders. The melting transitions of these rung-trimerized density waves belong to the same universality classes as those of density waves in the one-dimensional Rydberg chain and in two-leg Rydberg ladders. Upon further increasing the detuning, we observe an entanglement-entropy peak within the rung-trimerized density-wave phase, indicating a crossover regime with enhanced period-2 density modulation, followed by a first-order transition into long-range ordered states with one Rydberg excitation per occupied triangle that break the $\mathbb{D}_3$ symmetry. We also identify rung-trimerized floating phases between trimer density waves with different periods. For smaller blockade radii, where single occupation per triangle is allowed, the disordered phase first crosses over into a rung-trimerized phase without symmetry breaking and subsequently undergoes a first-order transition into a $\mathbb{D}_3$ symmetry-breaking phase. Finally, when the blockade radius is moderately larger than the lattice spacing, an extended critical region emerges between the disordered phase and a density-wave phase with alternating double and single Rydberg occupancy; both phase boundaries are of BKT type \cite{Berezinsky:1970fr,Kosterlitz_1973,Kosterlitz_1974}, capturing the low-energy physics of a $\mathbb{Z}_6$ clock model \cite{ORTIZ2012780}.

This paper is organized as follows. In Sec.~\ref{sec:model}, we introduce the Hamiltonian of the Rydberg triangular prism and summarize the numerical methods and diagnostics used to characterize the quantum phases. In Sec.~\ref{subsec:phasediagram}, we present the global phase diagrams and discuss the physical origin of the various phases and phase transitions. In Sec.~\ref{subsec:qtransitions}, we provide numerical evidence for the different quantum phase transitions and determine the corresponding transition points, critical exponents, and central charges. In Sec.~\ref{subsec:crossover}, we discuss the crossover regime inside the $\mathbb{Z}_p^+$ crystalline phases and present numerical evidence for its aspect-ratio dependence. Finally, in Sec.~\ref{sec:conclusion}, we summarize our main results and discuss possible future directions.

%%%%%%%%%%%%%%%%%%%%%%%%%%%%%%%%%%%%%%%%%%%%%%%%%%%%%%%%%%%%%%%%%%%%%%%%%%
\section{Model and Methods}\label{sec:model}
%%%%%%%%%%%%%%%%%%%%%%%%%%%%%%%%%%%%%%%%%%%%%%%%%%%%%%%%%%%%%%%%%%%%%%%%%%

%%%%%%%%%%%%%%%%%%%%%%%%%%%%%%%%%%%%%%%%%%%%%%%%%%%%%

\subsection{Hamiltonian}

We investigate a Rydberg atom array arranged in a triangular-prism geometry, consisting of three legs of one-dimensional (1D) chains whose cross sections form equilateral triangles. The side length of each triangle is set equal to the lattice spacing $a$ between neighboring triangular units. The system is governed by the Hamiltonian
\begin{eqnarray}\label{eq:rydbergham}
\nonumber \hat{H} &=& \sum_{i=1}^{L} \sum_{s = 1}^{3} \left( \frac{\Omega}{2}  \ket{g_{i,s}}\bra{r_{i,s}} + \text{h.c.} - \Delta \hat{n}_{i,s} \right)  \\
&+&  \sum_{\mathbf{r} \neq {\mathbf{r'}}} V_{\mathbf{r}, \mathbf{r'}} \hat{n}_{\mathbf{r}}\hat{n}_{\mathbf{r'}},
\end{eqnarray}
where $\Omega$ is the Rabi frequency, $\Delta$ is the laser detuning, and $V_{\mathbf{r}, \mathbf{r'}} = C_6 / |\mathbf{r} - \mathbf{r'}|^6$ denotes the van der Waals interaction. The Rydberg blockade radius $R_b$ is defined by equating the interaction strength at distance $R_b$ to the Rabi frequency, giving $R_b = (C_6/\Omega)^{1/6}$. In practice, we fix the energy scale by setting $\Omega=1$. Measuring length in units of the lattice spacing $a$, we have $R_b/a = V_0^{1/6}$, where $V_0$ is the interaction strength between nearest-neighbor (NN) sites. We therefore present the phase diagram in the $\Delta/\Omega$--$R_b/a$ plane.

In the regime of large $R_b/a$, the strong inter-leg repulsion energetically forbids multiple simultaneous excitations within the same triangular unit. The Hilbert space can then be projected onto an effective subspace in which each triangular unit cell contains at most one Rydberg excitation, $\sum_{s} \hat{n}_{i,s} \leq 1$. The resulting effective Hamiltonian is
\begin{equation}
\label{eq:rydberghameff}
\begin{aligned}
\hat{H}_{\rm eff}
&= \sum_{i=1}^{L}
\left[
\frac{\Omega}{2}\left( \hat{A}_{{\rm eff},i} + \hat{A}_{{\rm eff},i}^{\dagger} \right)
- \Delta \hat{N}_{i}
\right] \\
&+ \sum_{l=1}^{l_{\rm max}} \sum_{i=1}^{L-l} V_{l} \hat{N}_{i} \hat{N}_{i+l} + \sum_{l=1}^{l_{\rm max}} \sum_{i=1}^{L-l} \sum_{s=1}^3 U_{l}
\hat{D}_{s,i} \hat{D}_{s,i+l}.
\end{aligned}
\end{equation}
where $V_l = V_0[1/(4l^6)+3/(4(l^2+1)^3)]$ and $U_l=V_0[1/(4l^6)-1/(4(l^2+1)^3)]$. The operator $\hat{A}_{\rm eff}$ is the effective annihilation operator satisfying $\hat{A}_{\rm eff} \ket{100} = \hat{A}_{\rm eff} \ket{010} = \hat{A}_{\rm eff} \ket{001} = \ket{000}$, and $\hat{N}_{i}$ is the total Rydberg density in the $i$th triangle. Here $\hat{D}_{s,i}$ is a diagonal flavor-sign operator acting within the constrained local Hilbert space. It assigns eigenvalue $-1$ to the basis state with a Rydberg excitation on leg $s$, and eigenvalue $+1$ to the basis states with an excitation on the other two legs. Explicitly, $\hat{D}_{s,i} \hat{n}_{i,s} = - \hat{n}_{i,s}$ and $\hat{D}_{s,i} \hat{n}_{i,s'} = \hat{n}_{i,s'}$ for $s \neq s'$. The $V_l$ terms control the modulation of the total density, while the $U_l$ terms act as effective antiferromagnetic couplings that favor two occupied triangles having their Rydberg excitations on different legs. In both the original Hamiltonian and the effective Hamiltonian, we truncate the long-range interactions at $l_{\rm max}=20$.

\subsection{Methods}

We compute the von Neumann entanglement entropy $\mathcal{S}_{\rm vN}$ to map out the ground-state phase diagram. For a quasi-1D system of length $L$, the entanglement entropy of a subsystem $\mathcal{A}$ containing $l$ consecutive unit cells is defined as $\mathcal{S}_{\mathrm{vN}}(l) = -\operatorname{Tr}\,\rho_{\mathcal{A}}\ln \rho_{\mathcal{A}}$, where $\rho_{\mathcal{A}} = \Tr_{\bar{\mathcal{A}}} \ket{\Psi}\bra{\Psi}$ is the reduced density matrix obtained by tracing out the complement $\bar{\mathcal{A}}$, and $\ket{\Psi}$ is the ground state. At a $(1+1)$-dimensional conformal critical point with open boundary conditions (OBCs), the entanglement entropy of the leftmost $l$ consecutive unit cells grows logarithmically with subsystem size and follows the CFT scaling form \cite{HOLZHEY1994443,VidalEntangle2003,PasqualeCalabrese_2004,Calabrese_2009}
\begin{equation}
\label{eq:cfteeform}
\mathcal{S}_{\mathrm{vN}}= \frac{c}{6} \ln \left(\frac{4\left(L+1\right)}{\pi} \sin \left[\frac{\pi\left(2 l+1\right)}{2\left(L+1\right)}\right]\right) + s_o,
\end{equation}
where $c$ is the central charge and $s_o$ is a nonuniversal constant. By fitting the numerical data to Eq.~(\ref{eq:cfteeform}), we extract the central charge at critical points. In contrast, for gapped phases, $\mathcal{S}_{\rm vN}$ saturates to a finite value according to the area law \cite{RevModPhys.82.277} and is typically small unless the system lies close to a critical region. Therefore, low-entanglement regions generally correspond to gapped phases, enhanced-entanglement lines often indicate continuous phase boundaries, and extended high-entanglement regions signal gapless critical phases.

(Quasi-)long-range density-wave orders can be diagnosed from the local density profiles $\langle \hat{n}_{i,s} \rangle$ and the local trimer observable $\langle \hat{T}_i \rangle$, where
\begin{eqnarray}
\hat{T}_i = \sum_{s\neq s' \neq s''} \hat{a}^\dagger_{i,s} \hat{a}_{i,s'} \left(1-\hat{n}_{i,s''} \right).    
\end{eqnarray}
For the $\mathbb{Z}_p^+$ phase, we define the trimer-density-wave order parameter
\begin{eqnarray}
\hat{M}_p = \frac{1}{L} \sum_j \cos\!\left[\frac{2\pi}{p}(j-1)\right] \hat{T}_j,
\end{eqnarray}
which is finite in the $\mathbb{Z}_p^+$ phase and vanishes in the disordered phase. The continuous quantum phase transition between the disordered phase and the $\mathbb{Z}_p^+$ phase can be characterized by the Binder cumulant \cite{BinderMonte1985JoCP}
\begin{eqnarray}
U_4=1-\frac{\left\langle\hat{M}_p^4\right\rangle}{3\left\langle\hat{M}_p^2\right\rangle^2},
\end{eqnarray}
which is a dimensionless quantity exhibiting a common crossing point for different system sizes at the critical point. Fixing $R_b/a$ and tuning $\Delta/\Omega$ across the transition, the Binder cumulant near criticality satisfies the scaling form
\begin{eqnarray}
\label{eq:binderscaling}
U_4=f\left[L^{1 / \nu}\left(\Delta / \Omega-(\Delta / \Omega)_c\right)\right],
\end{eqnarray}
where $(\Delta/\Omega)_c$ is the critical point, $\nu$ is the correlation-length exponent, and $f(x)$ is a universal function. Another useful diagnostic is the energy gap $\Delta E$ between the ground state and the first excited state, which near the critical point obeys
\begin{eqnarray}
\label{eq:gapscaling}
L^z \Delta E=g\left[L^{1 / \nu}\left(\Delta / \Omega-(\Delta / \Omega)_c\right)\right],
\end{eqnarray}
where $z$ is the dynamical exponent and $g(x)$ is another universal function. Since $L^z\Delta E$ is also dimensionless, it likewise exhibits crossings for different system sizes. Based on the scaling forms in Eqs.~\eqref{eq:binderscaling} and \eqref{eq:gapscaling}, we perform data collapse for different system sizes near the critical point to extract $(\Delta/\Omega)_c$ as well as the critical exponents $\nu$ and $z$ \cite{ZhangFloating2025NC,LiaoPhaseRydLadder2025prb}. To locate the transition more accurately, we also determine the crossing points for pairs of system sizes with a fixed size increment and extrapolate these finite-size crossings to the thermodynamic limit. This procedure is particularly useful for BKT transitions, where logarithmic corrections lead to slowly converging crossings.

To further diagnose density modulations, we compute the structure factor, i.e., the Fourier transform of the density-density correlation function. When focusing on translational density oscillations, we use the structure factor of the total density in each triangle,
\begin{eqnarray}
\label{eq:strucN}
S_N(k) &=& \frac{1}{L^2} \sum_{l, l'} e^{ik(l - l')} \langle \hat{N}_{l} \hat{N}_{l'} \rangle,
\end{eqnarray}
where $k \in [0, 2\pi)$ is the wave vector. For phases with (quasi-)long-range order, and even for disordered phases with a moderate correlation length, peaks in $S_N(k)$ accurately identify the oscillation wave vector of the density-density correlations \cite{ZhangFloating2025NC,LiaoPhaseRydLadder2025prb}. In this way, we determine the loci of fixed wave vector in the phase diagram and track how the disordered phase evolves into phases with (quasi-)long-range order.

To characterize the critical phase, we extract the Luttinger parameter $K$ using the eigenstate crosscap overlap (ESCO) method introduced in Ref.~\cite{TanExtracting2025PRL}. For a one-component Tomonaga--Luttinger liquid with periodic boundary conditions (PBCs), conformal field theory predicts that the overlap between a crosscap state and certain low-lying states is a universal number determined solely by $K$. In the lattice formulation, the corresponding crosscap overlap is obtained by pairing antipodal unit cells into maximally entangled pairs. In our actual calculation, we do not explicitly construct the lattice crosscap state. Instead, we evaluate the equivalent quantity directly at the MPS level by contracting the physical index of the $i$th site with that of the $(i+L/2)$th site for all $i$, while leaving the virtual bond indices contracted in the standard MPS network. Denoting the resulting overlap by $X_{\rm cc}$, we extract $K$ from
\begin{equation}
\label{eq:crosscapK}
|X_{\rm cc}|^2 = K^{-1/2},
\end{equation}
that is, $K = |X_{\rm cc}|^{-4}$. We use the resulting $K$ to track the BKT transitions at the boundaries of the critical phase.

\subsection{Parameters of DMRG algorithms}

To obtain the ground states, we employ finite-size density matrix renormalization group (DMRG) \cite{DMRG1992prl,DMRG1993prb} calculations within the matrix product state (MPS) framework \cite{PhysRevLett.75.3537}. Our implementation is based on the \textsc{ITensor Julia} library \cite{10.21468/SciPostPhysCodeb.4}. To faithfully model the experimental setup, we retain all Rydberg interactions between sites belonging to triangles separated by at most 20 unit cells, i.e., within any block of 21 consecutive triangles. During the variational sweeps, we gradually increase the maximum bond dimension $D$ until the truncation error $\varepsilon$ is reduced to at least $10^{-10}$. In calculations of critical properties, we use smaller truncation errors when necessary, as specified explicitly in the corresponding discussions. We regard the calculation as converged when, over the final two sweeps, the changes in the ground-state energy and the von Neumann entanglement entropy are smaller than $10^{-11}$ and $10^{-8}$, respectively. While crystalline phases typically converge within a few dozen sweeps, the quantum floating phase may require thousands of sweeps, especially at larger Rydberg blockade radius $R_b/a$.

%%%%%%%%%%%%%%%%%%%%%%%%%%%%%%%%%%%%%%%%%%%%%%%%%%%%
\section{Results} \label{subsec:modelgsfs}
%%%%%%%%%%%%%%%%%%%%%%%%%%%%%%%%%%%%%%%%%%%%%%%%%%%%

\begin{figure*}[t]
\centering 
\includegraphics[width=1\textwidth]{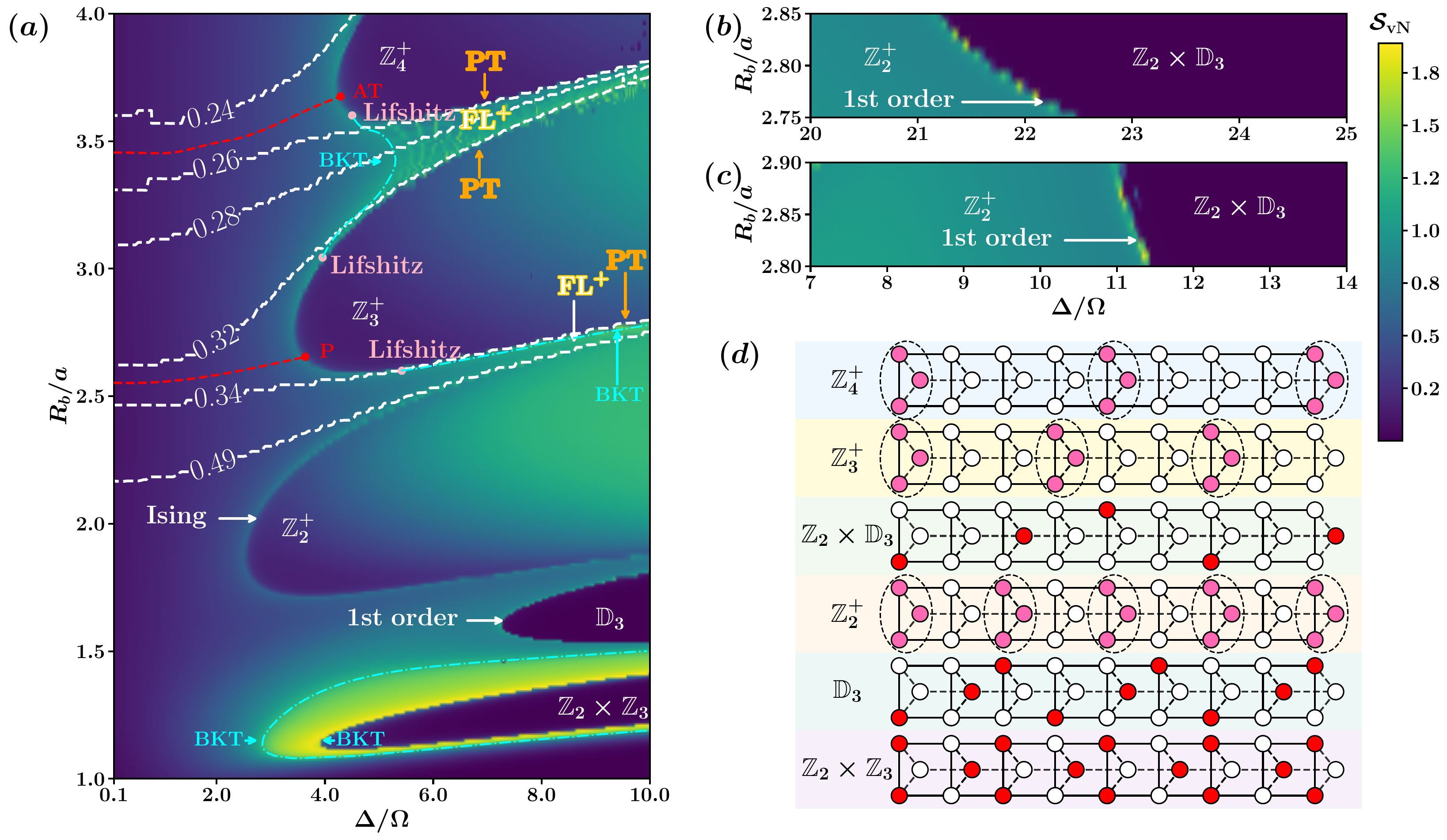}
\caption{(a) Ground-state phase diagram of the triangular-prism Rydberg atom array, obtained from the von Neumann entanglement entropy $\mathcal{S}_{\rm vN}$. The results are for $L=193$ triangles with OBCs, and $\mathcal{S}_{\rm vN}$ is evaluated across the cut between the 96th and 97th triangles. We use the original Hamiltonian in Eq.~\eqref{eq:rydbergham} for $R_b/a < 2.95$, and the effective Hamiltonian in Eq.~\eqref{eq:rydberghameff} for $R_b/a \ge 2.95$. The dark lobes denote distinct crystalline orders, whose density patterns are shown in (d). The dashed lines in the disordered region indicate constant-$k$ lines, where $k$ is the oscillation wave vector in the short-range density-density correlations; the labels on these lines give $k/2\pi$. There is an extended $k=\pi$ regime below constant $k/2\pi=0.49$ line. The $\mathbb{Z}_2\times\mathbb{Z}_3$ phase melts into the disordered phase through an intermediate six-state clock critical phase, with both transitions of BKT type. The $\mathbb{D}_3$ phase enters the disordered phase through a first-order transition. The $\mathbb{Z}_2^+$ phase melts through an Ising critical line. Along the commensurate lines, the $\mathbb{Z}_3^+$ and $\mathbb{Z}_4^+$ phases can transition directly into the disordered phase through conformal critical points, namely the Potts (P) and Ashkin-Teller (AT) points, respectively. Away from the CFT points along the phase boundaries of the corresponding lobes, the direct transitions are chiral. Further away from the CFT points, on both the smaller-$R_b/a$ and larger-$R_b/a$ sides, the direct boundary between the disordered phase and the crystalline order terminates at Lifshitz points, beyond which an incommensurate floating phase (FL$^+$) opens between the disordered phase and the crystalline order. The transition from the disordered phase to the floating phase is of BKT type, whereas that from the floating phase to the crystalline phases is of Pokrovsky-Talapov (PT) type.
(b) Enlarged phase diagram inside the $\mathbb{Z}_2^+$ lobe at large $\Delta/\Omega$, showing an additional first-order transition associated with spontaneous breaking of the $\mathbb{D}_3$ symmetry of the triangular cross section.
(c) Same as (b), but for the inter-triangle spacing $a_x=2a/3$, where $a$ is the side length of each triangle.
(d) Density patterns of the crystalline orders appearing in the phase diagram. Red filled circles denote sites with high Rydberg density (close to 1), while empty circles denote sites with low Rydberg density (close to 0). Pink filled circles in the circled triangles denote trimerized states of the form $(\ket{rgg}+\ket{grg}+\ket{ggr})/\sqrt{3}$.
}  
\label{fig:phasediagram}
\end{figure*}

In this section, we discuss the ground-state phase diagram and quantum phase transitions of the Rydberg triangular-prism system. The phase diagrams of Rydberg ladders with different aspect ratios have been studied extensively \cite{SarkarOrderbyDis2023,eck2023critical,JinPRDCritical2024,ZhangFloating2025NC,LiaoPhaseRydLadder2025prb,ChepigaLadder2025prr}. Although some transitions in the prism geometry fall into the same universality classes as in the ladder case, for example those between the $\mathbb{Z}_{p}^{+}$ ordered phases and the disordered phase, the higher symmetry of the triangular cross section leads to additional quantum phases and more varied quantum phase transitions.

\subsection{Phase diagram}
\label{subsec:phasediagram}

Figure~\ref{fig:phasediagram}(a) summarizes the phase structure of the triangular-prism array. As a convenient global diagnostic in the $\Delta/\Omega$--$R_b/a$ plane, we plot the bipartite entanglement entropy across the cut between triangles $96$ and $97$ for a system of $193$ triangles. Similar to the cases of Rydberg chains and ladders \cite{MaceiraChiral2022prr,ZhangFloating2025NC,LiaoPhaseRydLadder2025prb}, the small-$\Delta/\Omega$ region is in a disordered phase. In this regime, the Rydberg density is low, and the repulsive interaction energy is therefore too weak to induce spontaneous symmetry breaking.

Although the disordered phase has no long-range order, the repulsive interaction still induces short-range oscillations in the density-density correlations. To characterize these oscillations, we determine the peak position of the total-density structure factor $S_N(k)$ defined in Eq.~\eqref{eq:strucN}, and map out the corresponding constant-$k$ lines in the phase diagram \cite{LiaoPhaseRydLadder2025prb}. The resulting lines show that the dominant short-range density modulation depends sensitively on the parameters. In particular, there is an extended region with $k=\pi$ at small $R_b/a$, indicating that a period-2 density modulation is already favored by relatively weak repulsive interactions. For larger $R_b/a$, correlations with longer characteristic periods become increasingly important, and the extracted wave vector $k$ evolves continuously with the parameters. Consequently, while the $k=\pi$ regime occupies a finite region in parameter space, the commensurate loci with $k=2\pi/p$ for integers $p\ge 3$ appear only as isolated lines rather than extended regions.

At sufficiently large detuning, the peaks in $\mathcal{S}_{\rm vN}$ outline several low-entanglement lobes, which correspond to crystalline phases with spontaneous symmetry breaking. Three types of crystalline orders appear in the phase diagram. The first breaks only translational symmetry while preserving the $\mathbb{D}_3$ symmetry of the triangular cross section, and is labeled $\mathbb{Z}_p^+$. The second breaks translational symmetry together with the rotational $\mathbb{Z}_3$ symmetry of the triangle while preserving one reflection symmetry, and is labeled $\mathbb{Z}_p\times \mathbb{Z}_3$. The third breaks both translational symmetry and the full $\mathbb{D}_3$ symmetry of the triangle, and is labeled $\mathbb{Z}_p\times \mathbb{D}_3$. The corresponding density patterns are illustrated in Fig.~\ref{fig:phasediagram}(d). In the $\mathbb{Z}_p^+$ phases, the $\mathbb{D}_3$ symmetry is preserved by forming a trimerized state of the form $(\ket{rgg}+\ket{grg}+\ket{ggr})/\sqrt{3}$ on each occupied triangle, so that each site on such a triangle has an average Rydberg density close to $1/3$. This is analogous to the dimerized states in the ladder geometry, where the leg-exchange symmetry is preserved by symmetric bonding on each occupied rung \cite{eck2023critical,LiaoPhaseRydLadder2025prb}. Neighboring occupied triangles are separated by $p-1$ empty triangles, leading to translational symmetry breaking with period $p$. By contrast, in the other crystalline phases the occupied sites have Rydberg density close to $1$, and no trimerized or dimerized superposition is formed. As a result, the $\mathbb{D}_3$ symmetry of the triangular cross section is broken either partially or completely.

For $R_b/a$ slightly above $1$, large detuning stabilizes a $\mathbb{Z}_2\times \mathbb{Z}_3$ order. This phase breaks translation symmetry along the prism axis down to a period-two density pattern, in which a triangle with two occupied sites alternates with a neighboring triangle with a single occupied site on the third vertex. Starting from one such configuration, the other five symmetry-related configurations are generated by translation along the prism axis and cyclic permutation of the three legs, yielding a sixfold ground-state degeneracy. In this phase, the $\mathbb{D}_3$ symmetry of the triangular cross section is only partially broken, since one reflection symmetry, namely that about the perpendicular bisector of the two occupied sites, remains preserved. Because only the $\mathbb{Z}_2$ translational symmetry is broken, domain-wall excitations do not generate chiral perturbations \cite{HuseFisherPRL1982}. The melting of the $\mathbb{Z}_2\times \mathbb{Z}_3$ order is therefore a commensurate-to-commensurate transition, consistent with the fact that the adjacent disordered regime in this range of $R_b/a$ is characterized by commensurate density oscillations with $k=\pi$. As a result, the $\mathbb{Z}_2\times \mathbb{Z}_3$ phase first enters an intermediate critical phase through a BKT transition, and this critical phase then crosses into the disordered phase through a second BKT transition. This gives rise to an extended regime that realizes the physics of the six-state clock model. This behavior is in contrast to crystalline phases with higher-period translational symmetry breaking, where the two-step melting instead proceeds through an incommensurate floating phase bounded by a PT transition on the ordered side \cite{Pokrovsky1979}.

With increasing $R_b/a$, additional crystalline phases appear. The first one is the $\mathbb{D}_3$ phase, which is favored once the blockade radius becomes large enough that no more than one Rydberg excitation can reside on a triangle. In this regime, NN sites on the same leg cannot be occupied simultaneously, and the repulsion between next-nearest-neighbor (NNN) sites along the same leg further penalizes configurations with nearby occupations on the same leg. As a result, the interaction energy is minimized by a pattern in which three consecutive triangles each carry one excitation, with the occupied site rotating through the three legs from triangle to triangle. In this phase, each individual triangle breaks the rotational $\mathbb{Z}_3$ symmetry and preserves only one reflection symmetry. However, because the reflection axis changes from one triangle to the next along the winding pattern, the state as a whole fully breaks the $\mathbb{D}_3$ symmetry of the triangular cross section. It also breaks translational symmetry down to period $3$. Importantly, translation by one triangle is equivalent, within the ordered manifold, to a cyclic permutation of the three legs. Therefore, the six symmetry-related ground states do not form a $\mathbb{Z}_3$ translational sector together with an independent $\mathbb{Z}_2$ sector, but instead realize the nonabelian $\mathbb{D}_3$ structure itself. The three possible starting legs, combined with the two possible chiralities, give a sixfold degeneracy. This is fundamentally different from the $\mathbb{Z}_2\times\mathbb{Z}_3$ phase discussed above. In the latter case, translation and cyclic leg permutation commute and can be combined into a single order-6 operation, so the ordered manifold is naturally described by a $\mathbb{Z}_6$ clock variable. By contrast, no such order-6 generator exists for the $\mathbb{D}_3$ phase, since its six states are related by the noncommuting operations of the dihedral group. As discussed below, the $\mathbb{D}_3$ phase enters the disordered phase through a first-order transition.

For larger $R_b/a$, the phase diagram is dominated by the $\mathbb{Z}_p^+$ phases with $p=2,3,4$. In this regime, the strong repulsion suppresses double occupation of neighboring triangles, so the total density develops period-$p$ modulations along the prism axis. At the same time, in the moderate-detuning regime, the antiferromagnetic term in the effective Hamiltonian, Eq.~\eqref{eq:rydberghameff}, is too weak beyond NN triangles to induce spontaneous breaking of the $\mathbb{D}_3$ symmetry of the triangular cross section. As a result, each occupied triangle remains in a trimerized state of the form $(\ket{rgg}+\ket{grg}+\ket{ggr})/\sqrt{3}$, and only the translational $\mathbb{Z}_p$ symmetry is broken. The superscript ``$+$'' indicates that the $\mathbb{D}_3$ symmetry does not contribute an additional degeneracy factor. Consequently, the $\mathbb{Z}_p^+$ phases have a $p$-fold ground-state degeneracy associated solely with the broken translation symmetry. They are the natural prism analogs of the dimerized crystalline phases in the ladder geometry, and their physics is very similar to that of the corresponding ladder phases \cite{eck2023critical,ZhangFloating2025NC,LiaoPhaseRydLadder2025prb,ChepigaLadder2025prr}. In particular, the $\mathbb{Z}_2^+$ phase melts into the disordered phase through an Ising critical line. For the $\mathbb{Z}_3^+$ and $\mathbb{Z}_4^+$ phases, the direct transitions along the commensurate lines are controlled by the Potts and Ashkin-Teller (AT) critical points, respectively. Away from these CFT points along the same phase boundaries, the direct transitions are chiral with dynamical exponent $z>1$ \cite{WhitsittQuantum2018PRB}. Further away, on both the smaller-$R_b/a$ and larger-$R_b/a$ sides, the direct boundary between the disordered phase and the crystalline order terminates at Lifshitz points. Beyond these points, an incommensurate floating phase opens between the disordered phase and the commensurate ordered phase. The transition from the disordered phase to the floating phase is of BKT type, whereas that from the floating phase to the crystalline order is of PT type.

Another notable feature inside the $\mathbb{Z}_p^+$ lobes is a crossover regime, which is most clearly visible inside the $\mathbb{Z}_2^+$ phase in the parameter range studied here. As $\Delta/\Omega$ increases, the repulsion between occupied triangles becomes stronger, and the system tends to develop an additional period-2 density modulation. In the ladder geometry, this tendency is compatible with breaking the leg-exchange symmetry and can evolve directly into a $\mathbb{Z}_4$ order on each leg. In the prism geometry, however, such a tendency is incompatible with the eventual breaking of the $\mathbb{D}_3$ symmetry of the triangular cross section, which instead is associated with an additional period-3 modulation on each leg. As a result, the system first develops a pronounced tendency toward an additional $\mathbb{Z}_2$ modulation within the occupied-triangle sublattice, but this tendency does not become a true long-range order. Upon further increasing $\Delta/\Omega$, the system instead deviates from this tendency and enters a $\mathbb{Z}_p\times \mathbb{D}_3$ phase. Across this transition, the period-$p$ modulation of the total density is retained, while the breaking of the $\mathbb{D}_3$ symmetry generates an additional period-3 modulation on each leg, so that the density pattern along an individual leg has period $3p$. To illustrate this behavior, Fig.~\ref{fig:phasediagram}(b) shows the enlarged phase diagram for $20<\Delta/\Omega<25$ inside the $\mathbb{Z}_2^+$ lobe, where the transition to the $\mathbb{Z}_2\times \mathbb{D}_3$ phase is clearly seen. Figure~\ref{fig:phasediagram}(c) shows the corresponding phase diagram for a shorter inter-triangle spacing, $a_x=2a/3$, in the range $7<\Delta/\Omega<14$, where the same transition also appears. The first-order nature of this transition, as well as the microscopic structure of the crossover regime, will be discussed in the following subsections.

\subsection{Quantum phase transitions}
\label{subsec:qtransitions}
\subsubsection{Ising transition}

\begin{figure}[t]
    \centering
    \includegraphics[width=1\linewidth]{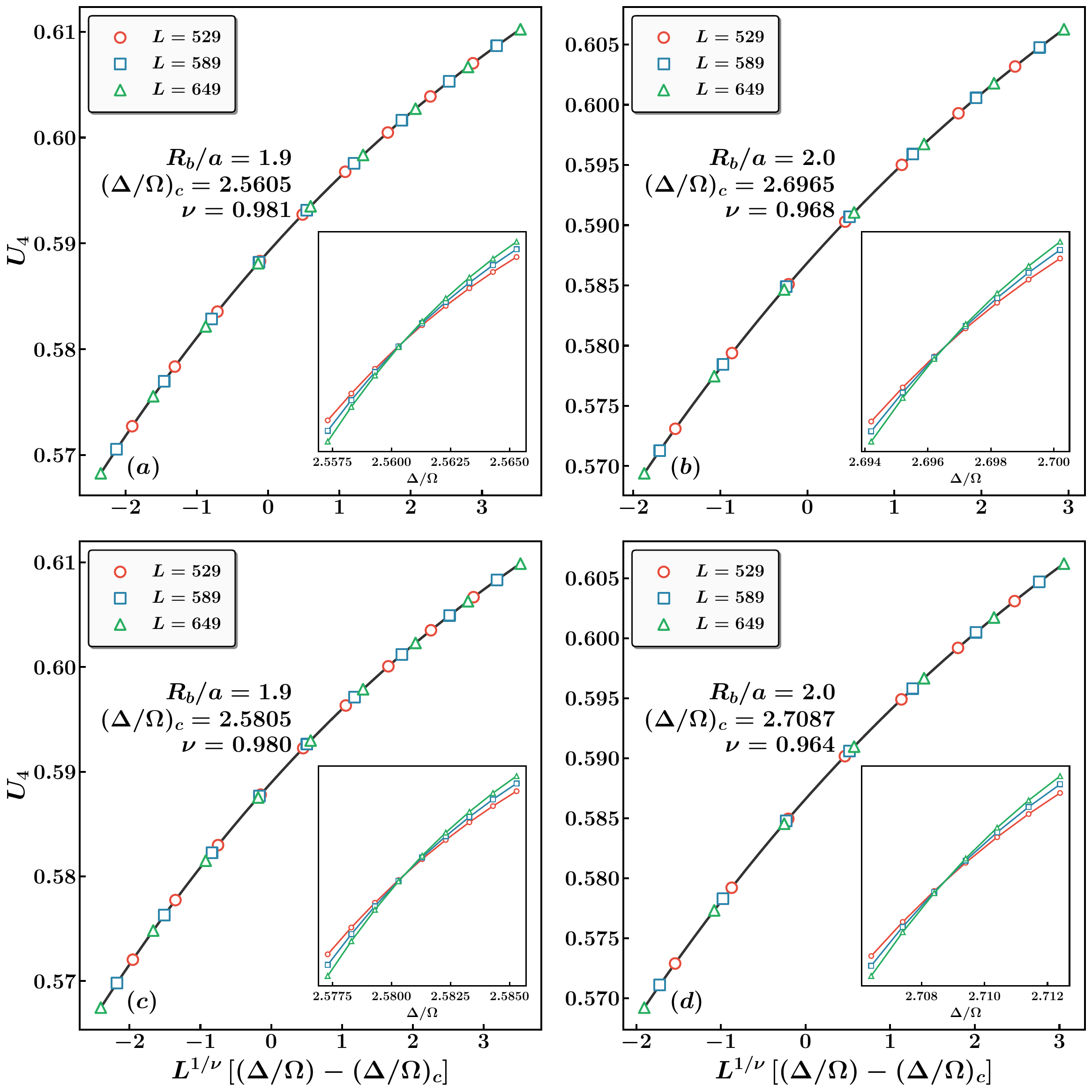}
    \caption{Data collapse of the Binder cumulant $U_4$ for the Ising transitions between the $\mathbb{Z}_{2}^{+}$ phase and the disordered phase. Panels (a) and (b) are along the cuts $R_b/a=1.9$ and $2.0$, respectively, calculated from the original Hamiltonian~\eqref{eq:rydbergham}. Panels (c) and (d) are for the same two cuts, but calculated from the effective Hamiltonian~\eqref{eq:rydberghameff}. The solid lines are polynomial fits to the collapsed data. The insets show $U_4$ as a function of $\Delta/\Omega$. The DMRG truncation error is set to $\varepsilon=10^{-12}$.}
    \label{fig:Ising}
\end{figure}

The transition from the disordered phase to the $\mathbb{Z}_{2}^{+}$ phase is associated with spontaneous breaking of translational $\mathbb{Z}_{2}$ symmetry, and is therefore expected to belong to the Ising universality class. This is also consistent with the presence of an extended commensurate regime with $k/2\pi=1/2$ in the phase diagram. To determine the critical points and the correlation-length exponent $\nu$, we compute the Binder cumulant $U_4$ of the order parameter $\hat{M}_{p=2}$ for system sizes $L=529$, $589$, and $649$. We then perform a data collapse within a narrow window of $\Delta/\Omega$ around the transition, with a width of about $0.01$ and a step size of $0.001$, by fitting $U_4$ as a function of $L^{1/\nu}[\Delta/\Omega-(\Delta/\Omega)_c]$ to a 9th-degree polynomial. The resulting collapse, shown in Fig.~\ref{fig:Ising}, gives $\nu=0.981$ and $0.980$ at $R_b/a=1.9$ for the original Hamiltonian in Eq.~\eqref{eq:rydbergham} and the effective Hamiltonian in Eq.~\eqref{eq:rydberghameff}, respectively. At $R_b/a=2.0$, the extracted values are $\nu=0.968$ and $0.964$, consistent with the Ising CFT prediction $c=1$. The agreement between the two Hamiltonians is therefore excellent, with differences of only $0.001$ and $0.004$ for the two cuts.

\subsubsection{$\mathbb{Z}_{3}^{+}$ order to the disordered phase}

\begin{figure}[t]
    \centering
    \includegraphics[width=1\linewidth]{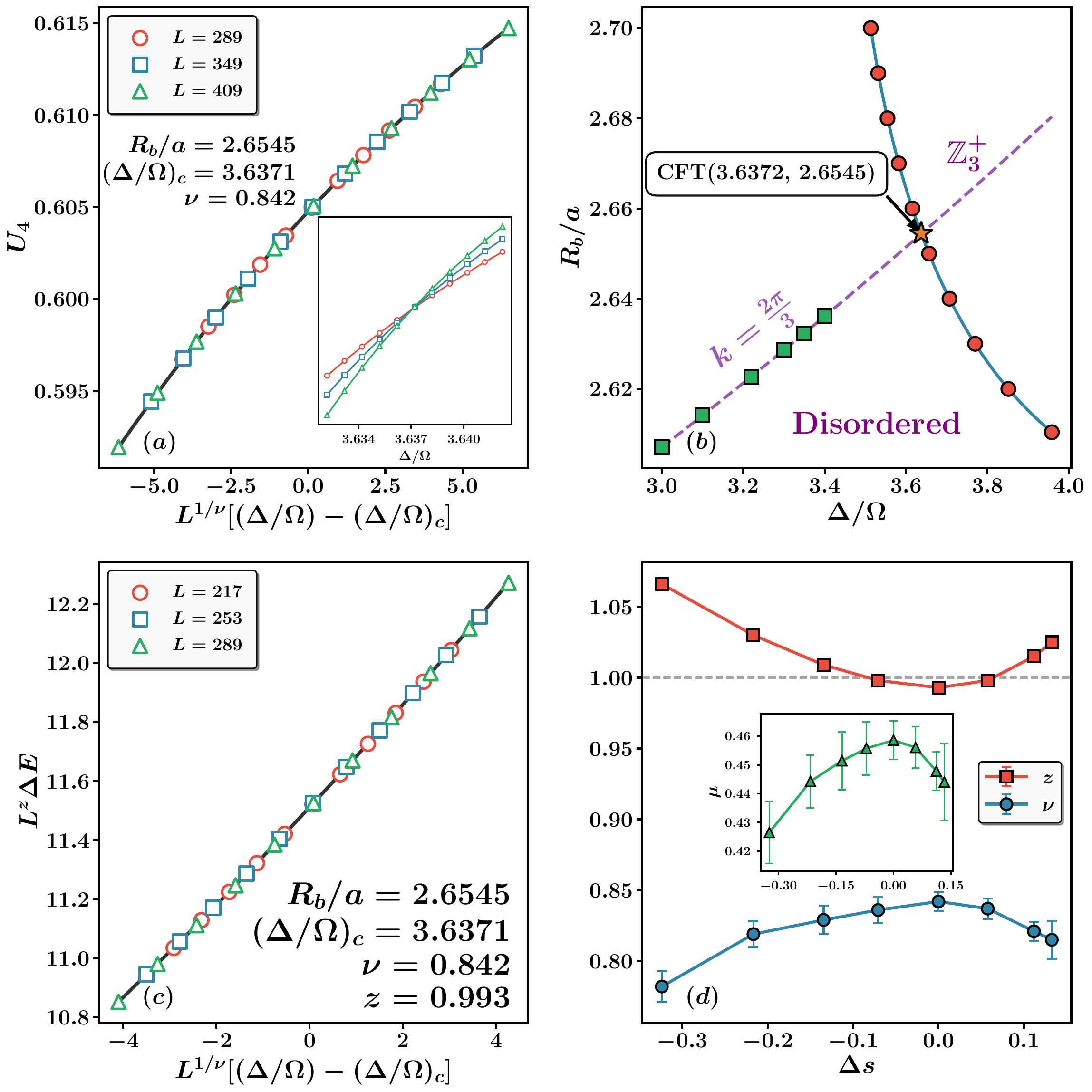}
    \caption{(a) Data collapse of the Binder cumulant $U_4$ for the quantum phase transition between the $\mathbb{Z}_{3}^{+}$ phase and the disordered phase at the Potts CFT point. (b) Phase boundary of the $\mathbb{Z}_{3}^{+}$ phase and the commensurate line with $k=2\pi/3$ in the disordered phase. Solid circles denote transition points determined from $U_4$ data collapse and are connected by spline interpolation. Diamonds denote extrapolated points at which the structure-factor peak is located at $k=2\pi/3$ in the thermodynamic limit. These points are connected by spline interpolation, and the resulting curve is extrapolated to larger $\Delta/\Omega$ until it intersects the phase boundary at the Potts CFT point. (c) Data collapse of the rescaled energy gap $L^z \Delta E$ at the CFT point on the $\mathbb{Z}_{3}^{+}$ boundary. (d) Correlation-length exponent $\nu$ and dynamical exponent $z$ for the transitions between the $\mathbb{Z}_{3}^{+}$ phase and the disordered phase in the vicinity of the CFT point. The horizontal axis denotes the signed displacement $\Delta_s$ from the CFT point along the phase boundary. The inset shows the Kibble-Zurek exponent $\mu=\nu/(1+z\nu)$. The DMRG truncation error is set to $\varepsilon=10^{-12}$ in (a) and $\varepsilon=10^{-10}$ in (c).}
    \label{fig:Z3+}
\end{figure}

The melting of the $\mathbb{Z}_3^+$ phase is mostly through commensurate-to-incommensurate chiral transitions, except at a special CFT point on the phase boundary. As shown in Fig.~\ref{fig:phasediagram}(a), this point is given by the intersection of the $\mathbb{Z}_3^+$ phase boundary and the commensurate line with $k=2\pi/3$ in the disordered phase. To locate it precisely, we first determine the $\mathbb{Z}_3^+$ boundary from Binder-cumulant data collapse for the order parameter $\hat{M}_{p=3}$, and then extrapolate the $k=2\pi/3$ commensurate line to the thermodynamic limit from the peak positions of the structure factor defined in Eq.~\eqref{eq:strucN}. The resulting phase boundary is shown in Fig.~\ref{fig:Z3+}(b), where ten critical points are obtained and connected by spline interpolation. The intersection with the commensurate line gives the 3-state Potts CFT point at $(\Delta/\Omega,R_b/a)=(3.6372,2.6545)$.

Figure~\ref{fig:Z3+}(a) shows the Binder cumulant along the cut $R_b/a=2.6545$, which passes through the Potts CFT point, for system sizes up to $L=409$. The optimal data collapse gives $(\Delta/\Omega)_c=3.6371$, consistent with the intersection point in Fig.~\ref{fig:Z3+}(b), and a correlation-length exponent $\nu=0.842$, which differs from the three-state Potts value $\nu=5/6$ by only about $1\%$. We further determine the dynamical exponent $z$ from the data collapse of the rescaled energy gap $L^z\Delta E$. A representative collapse at the Potts CFT point is shown in Fig.~\ref{fig:Z3+}(c) for $L=217$, $253$, and $289$. Fixing $(\Delta/\Omega)_c$ and $\nu$ from the Binder-cumulant analysis, we obtain the best collapse at $z=0.993$, in very good agreement with the CFT expectation $z=1$.

We plot the values of $\nu$ and $z$ as functions of the signed displacement $\Delta_s$ from the Potts CFT point along the phase boundary in Fig.~\ref{fig:Z3+}(d). As shown there, $\nu$ reaches a local maximum at the CFT point, while $z$ reaches a minimum there and then increases as one moves away from it. This is consistent with the behavior found in Rydberg chains and ladders \cite{PhysRevB.106.165124,ZhangFloating2025NC,LiaoPhaseRydLadder2025prb}. Correspondingly, the Kibble-Zurek exponent $\mu=\nu/(1+z\nu)$, shown in the inset, reaches a maximum of about $0.46$ at the CFT point. This gives a concrete experimental prediction for Rydberg-atom arrays. \cite{Keesling2019Kibble}. All these observations are consistent with the picture that, away from the CFT point, the direct transition becomes chiral and is characterized by $z>1$.

\subsubsection{$\mathbb{Z}_{4}^{+}$ order to the disordered phase}

The melting of the $\mathbb{Z}_4^+$ phase follows the same overall structure as that of the $\mathbb{Z}_3^+$ phase. Most of the phase boundary is chiral, while a single CFT point appears where the $\mathbb{Z}_4^+$ boundary intersects the commensurate line with $k=\pi/2$, as shown in Fig.~\ref{fig:phasediagram}(a). This point belongs to the AT universality class \cite{AshkinTeller1943}, for which the correlation-length exponent lies between the four-state Potts value $\nu=2/3$ and the four-state clock value $\nu=1$ \cite{Kohmoto1981AT}. In addition, chiral perturbations are relevant for $\nu\in(0.683,1)$ \cite{schulz1983phase}, where the direct transition is expected to become chiral immediately away from the AT point.

\begin{figure}[t]
    \centering
    \includegraphics[width=1\linewidth]{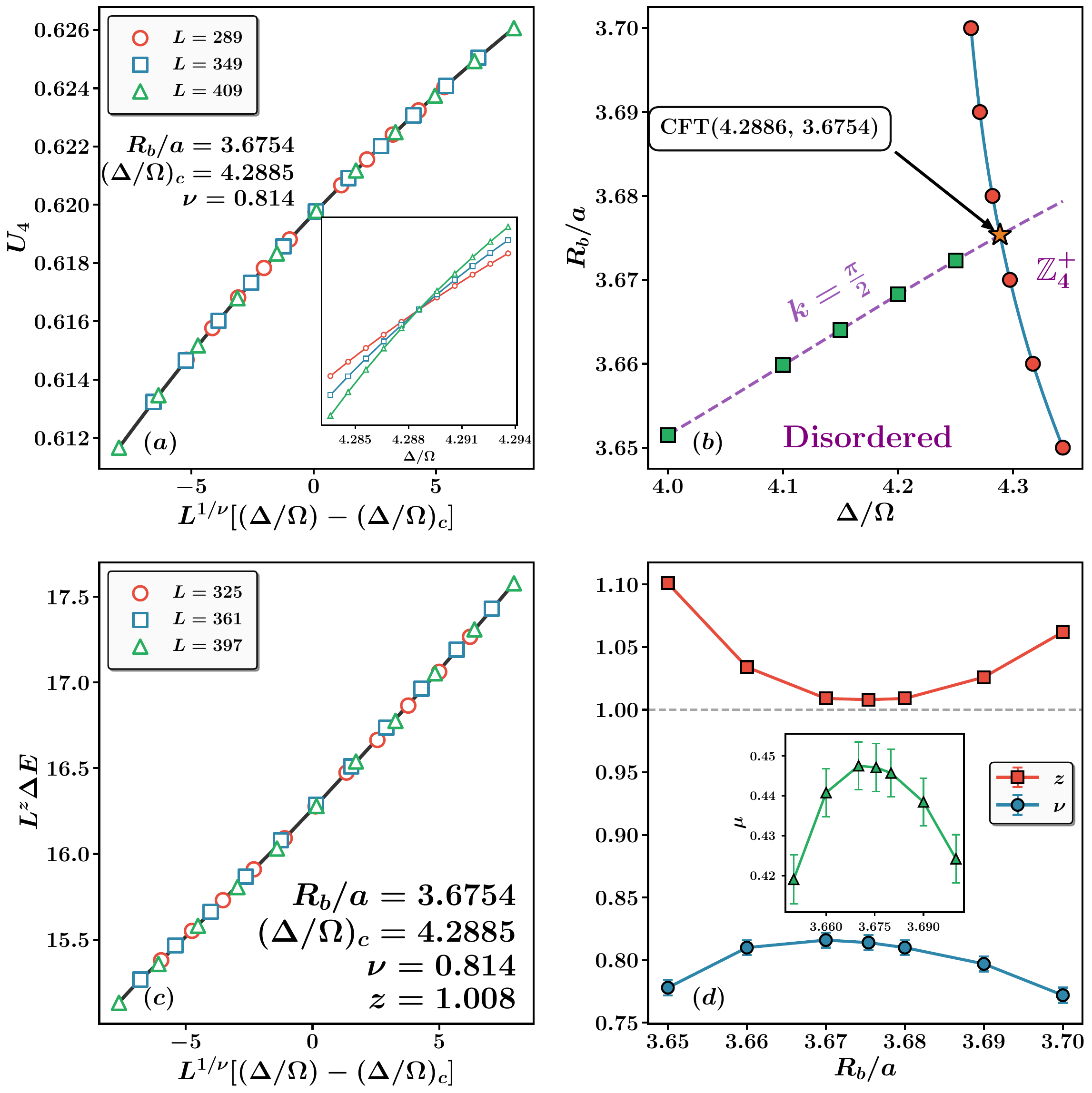}
    \caption{Same as Fig.~\ref{fig:Z3+}, but for the phase transition between the $\mathbb{Z}_{4}^{+}$ phase and the disordered phase. The commensurate line in (b) has wave vector $k=\pi/2$, and its intersection with the phase boundary determines the Ashkin-Teller CFT point. The DMRG truncation error is set to $\varepsilon=10^{-12}$ in (a) and $\varepsilon=10^{-10}$ in (c).}
    \label{fig:Z4+}
\end{figure}

Using the same procedure as for the $\mathbb{Z}_3^+$ phase, we determine the $\mathbb{Z}_4^+$ phase boundary from Binder-cumulant data collapse for $\hat{M}_{p=4}$ and the commensurate line with $k=\pi/2$ from the structure-factor peak positions. The resulting boundary is shown in Fig.~\ref{fig:Z4+}(b), where six critical points are obtained and connected by spline interpolation. Their intersection gives the CFT point at $(\Delta/\Omega,R_b/a)=(4.2886,3.6754)$. Figure~\ref{fig:Z4+}(a) shows the Binder cumulant along the cut $R_b/a=3.6754$ through this point. The optimal data collapse gives $(\Delta/\Omega)_c=4.2885$, consistent with the intersection value, and $\nu=0.815$. This supports the identification of this point as the unique AT CFT point on the $\mathbb{Z}_4^+$ boundary.

We then determine the dynamical exponent $z$ from the data collapse of the rescaled energy gap $L^z\Delta E$. As shown in Fig.~\ref{fig:Z4+}(c), the best collapse gives $z=1.008$, again in excellent agreement with the CFT expectation $z=1$. The critical exponents as functions of $R_b/a$ are summarized in Fig.~\ref{fig:Z4+}(d). As in the $\mathbb{Z}_3^+$ case, the CFT point corresponds to a local maximum of $\nu$ and a minimum of $z$. The corresponding Kibble-Zurek exponent also reaches a maximum near this point, with $\mu\approx 0.45$. The numerical observation of $z>1$ away from the CFT point shows that the direct transition becomes chiral.

\subsubsection{Melting of the $\mathbb{Z}_{2}\times\mathbb{Z}_{3}$ order}

\begin{figure}[b]
    \centering
    \includegraphics[width=1\linewidth]{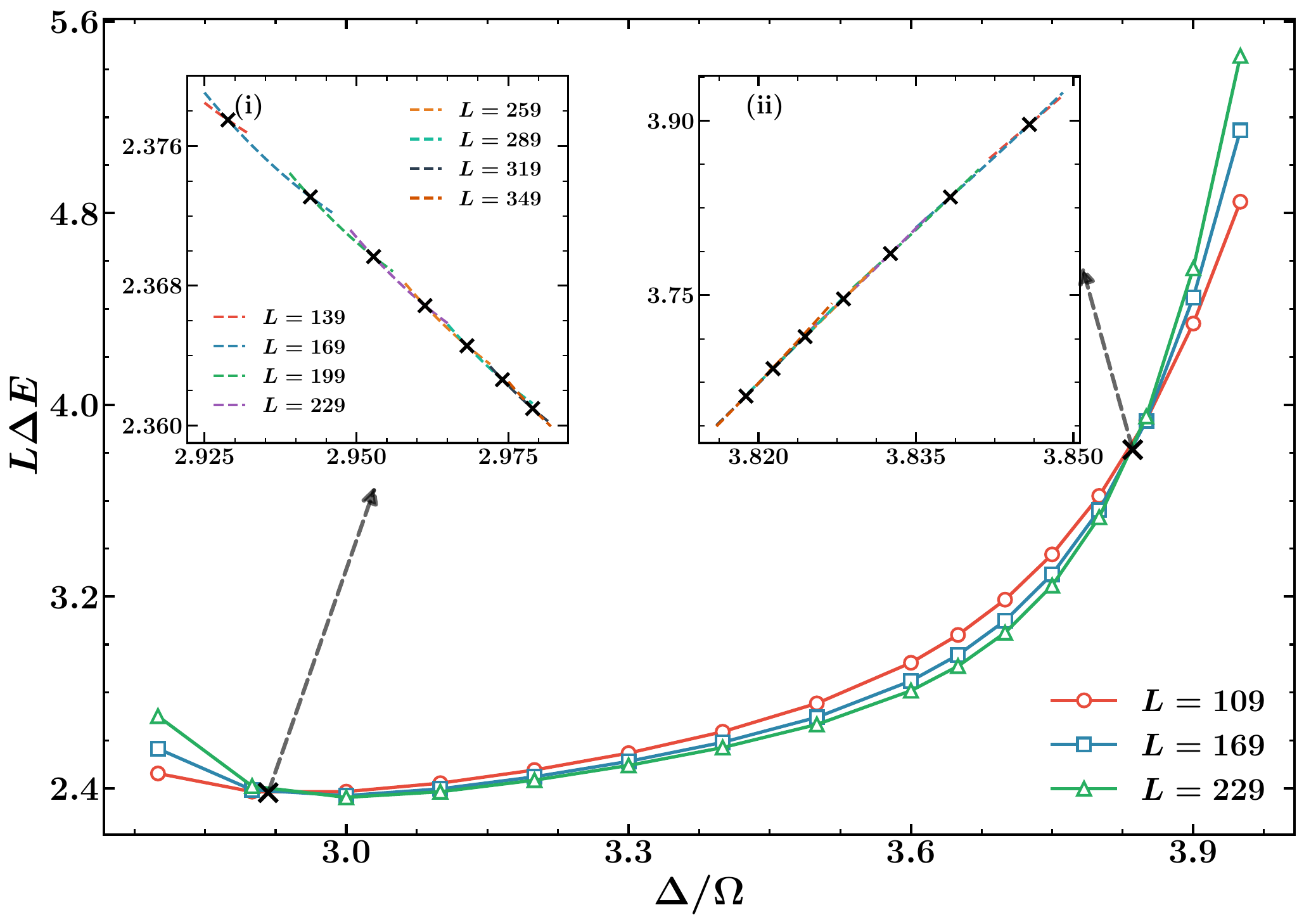}
    \caption{Rescaled energy gap $L\Delta E$ as a function of $\Delta/\Omega$ along the cut $R_b/a=1.14$, crossing from the disordered phase through the critical phase into the $\mathbb{Z}_2\times\mathbb{Z}_3$ ordered phase, for various system sizes $L$. Interactions are truncated at the fourth-nearest-neighbor triangles. The two crossing points indicate two BKT transitions. The insets show enlarged views of the two crossing regions, using crossings between system sizes $L$ and $L+30$ with $L=139,169,\ldots,319$, and illustrate the drift of the crossing points with increasing $L$.}
    \label{fig:Z2Z3_gap}
\end{figure}

\begin{figure}[t]
    \centering
    \includegraphics[width=1\linewidth]{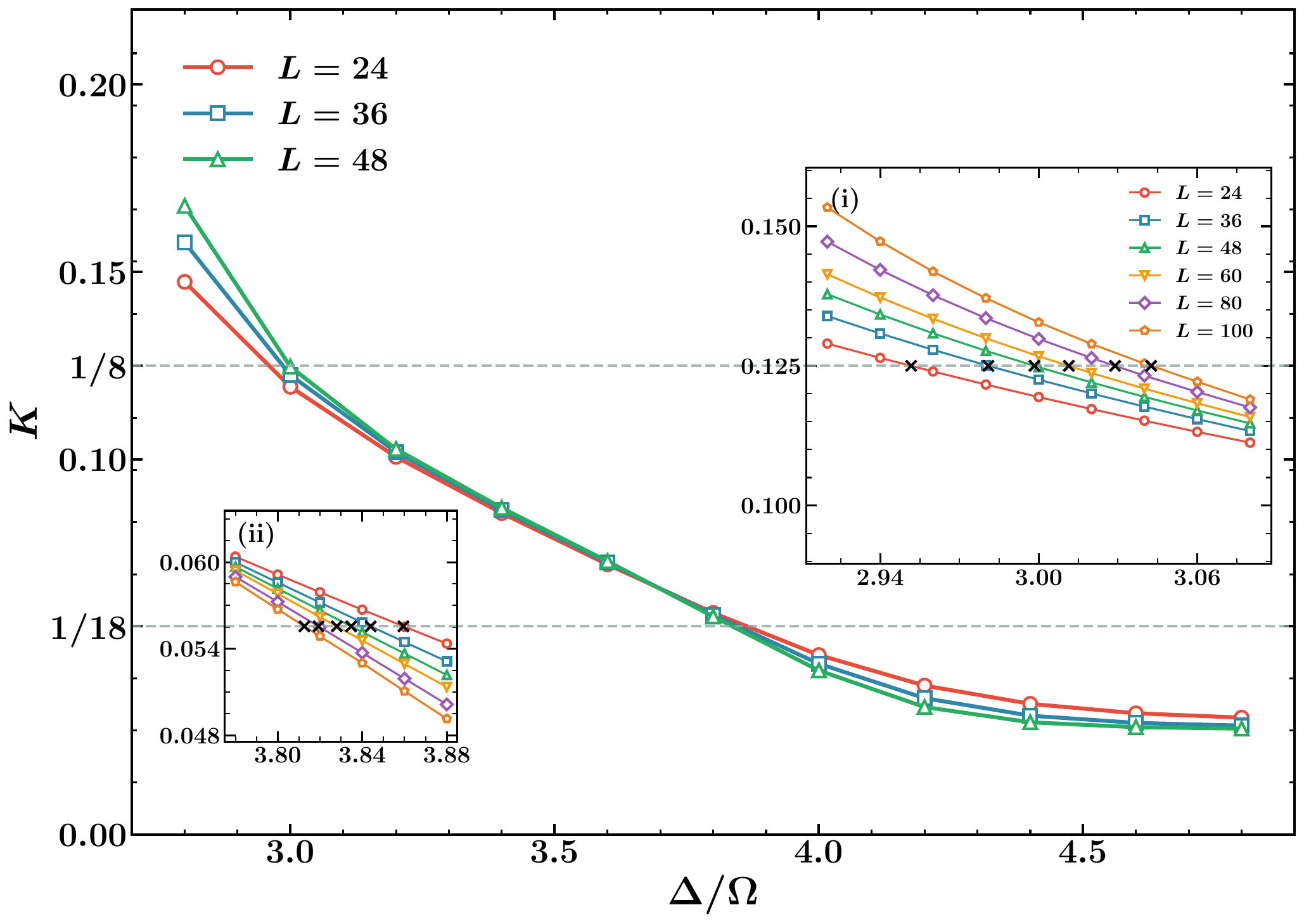}
    \caption{Luttinger parameter $K$ as a function of $\Delta/\Omega$ for various system sizes $L$, along the same cut as in Fig.~\ref{fig:Z2Z3_gap}, extracted from the crosscap overlaps under PBC. The same interaction truncation as in Fig.~\ref{fig:Z2Z3_gap} is used. The curves of $K$ intersect the two horizontal dashed lines at the critical values $K_c=1/18$ and $K_c=1/8$, which determine the two BKT transitions. The two insets show enlarged views near these two critical values using data for $L=24,36,48,60,72,84$, illustrating how the intersections with the critical values $K_c$ evolve with system size. The bond dimension is set to $D=600$.}
    \label{fig:Z2Z3_Luttinger}
\end{figure}

\begin{figure}[t]
    \centering
    \includegraphics[width=1\linewidth]{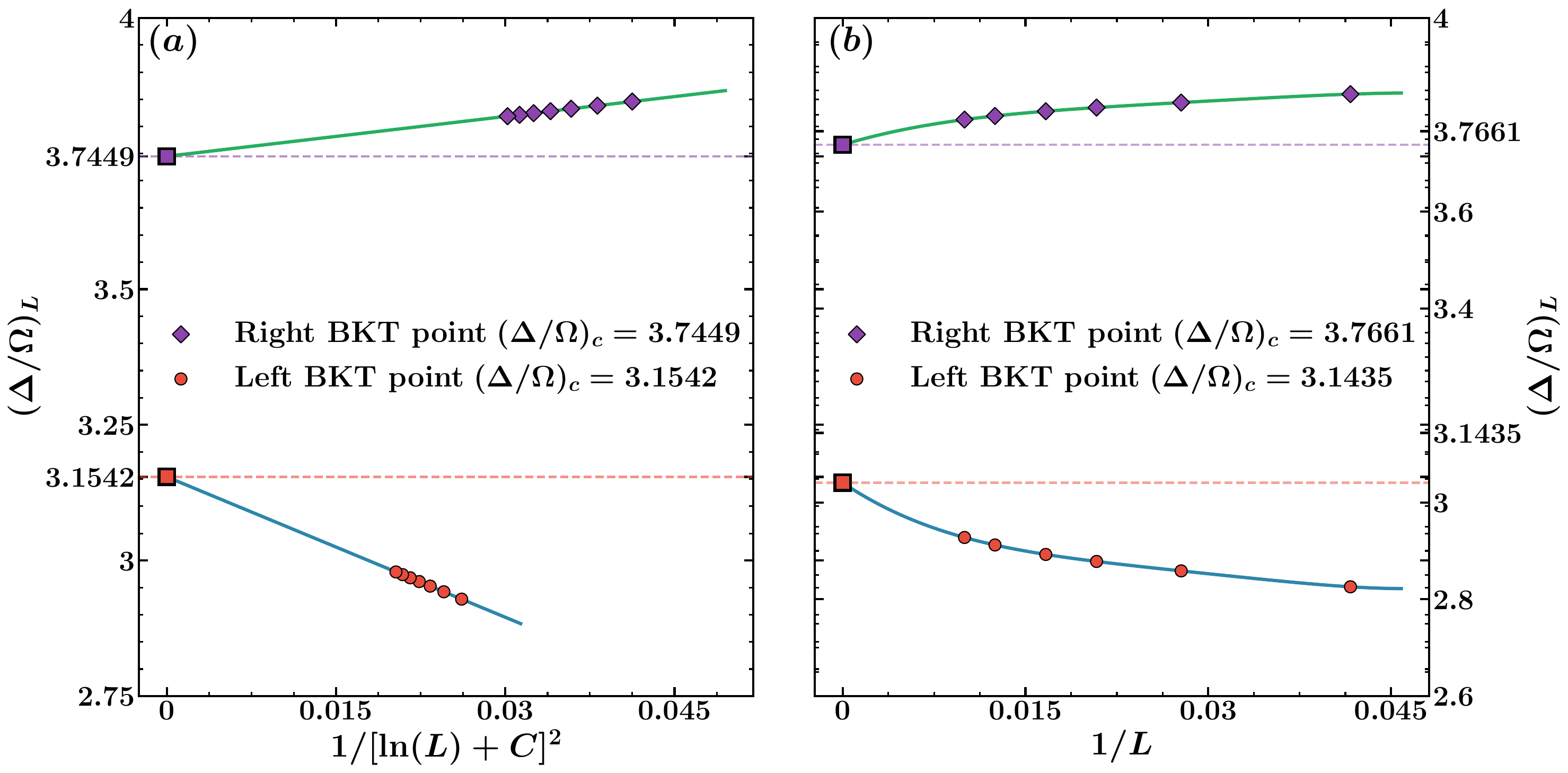}
    \caption{Extrapolation of the BKT transition points to the thermodynamic limit. (a) The finite-size crossing points of the rescaled energy gap in Fig.~\ref{fig:Z2Z3_gap} are fit to $A/[\ln(L)+C]^2+(\Delta/\Omega)_c$, where $(\Delta/\Omega)_c$ is the extrapolated BKT transition point and $A$ and $C$ are fitting parameters. (b) The finite-size locations of the intersections in Fig.~\ref{fig:Z2Z3_Luttinger} are fit to polynomials in $1/L$.}
    \label{fig:BKT_Luttinger_comparison}
\end{figure}
For the $\mathbb{Z}_2\times \mathbb{Z}_3$ ordered phase, the direct transition between the disordered phase and the ordered phase is replaced by an intermediate critical phase. As discussed in Sec.~\ref{subsec:phasediagram}, the low-energy manifold of the ordered phase is effectively described by a $\mathbb{Z}_6$ clock variable. Therefore, both the transition from the ordered phase to the critical phase and that from the critical phase to the disordered phase are expected to be of BKT type. Inside the intermediate critical phase, the low-energy physics is described by a Gaussian theory with a single Luttinger parameter $K$ \cite{GiamarchiQuantum2003}. The two BKT transitions are determined by the points where the symmetry-allowed perturbations become marginal. For the transition from the critical phase to the disordered phase, the relevant perturbation is the Rabi term, namely the creation and annihilation operator, whose scaling dimension is $1/(4K)$. Requiring this perturbation to be marginal gives the critical value $K_c=1/8$. For the transition from the critical phase to the $\mathbb{Z}_2\times\mathbb{Z}_3$ ordered phase, the relevant perturbation is the sixfold anisotropy associated with the effective $\mathbb{Z}_6$ clock order, which becomes marginal at $K_c=1/18$ \cite{GiamarchiQuantum2003}.

To locate these two transitions, we first examine the rescaled energy gap $L\Delta E$ as a function of the detuning $\Delta/\Omega$. As shown in Fig.~\ref{fig:Z2Z3_gap}, two distinct crossing points appear for different system sizes $L$, identifying the two boundaries of the intermediate critical phase. We then extrapolate the locations of these finite-size crossing points to the thermodynamic limit using the standard BKT scaling form $1/[\ln(L)+C]^2$. As shown in Fig.~\ref{fig:BKT_Luttinger_comparison}(a), the two BKT transition points are $(\Delta/\Omega)_c=3.1542$ and $3.7449$, respectively. We next determine the Luttinger parameter $K$ from Eq.~\eqref{eq:crosscapK} using the ESCO method under PBC \cite{TanExtracting2025PRL}. The results are shown in Fig.~\ref{fig:Z2Z3_Luttinger}. The two BKT transitions are identified by the intersections of the $K$ curves with the critical values $K_c=1/18$ and $K_c=1/8$. The corresponding thermodynamic extrapolation is shown in Fig.~\ref{fig:BKT_Luttinger_comparison}(b), giving $(\Delta/\Omega)_c=3.1435$ and $3.7661$, respectively. The critical points obtained from the gap crossings and from the Luttinger parameter agree very well, which further supports the $\mathbb{Z}_6$ clock description and the BKT nature of both transitions.

In both the rescaled-gap and ESCO analyses, the finite-size transition point $(\Delta/\Omega)_L$ increases with $L$ for the disordered-to-critical transition, but decreases with $L$ for the critical-to-ordered transition. This opposite drift can be understood from the different mechanisms that destabilize the critical phase on the two sides. On the ordered side, the transition is driven by domain-wall excitations, which are local defects and therefore weaken the ordered phase more strongly in smaller systems. As a result, a larger detuning is needed at small $L$ to stabilize the ordered phase, while for larger systems a smaller detuning is sufficient. On the disordered side, the destabilization of the critical phase is associated with long-wavelength collective fluctuations. Such nonlocal fluctuations are easier to accommodate in larger systems, so the critical phase is destroyed already at smaller Rabi frequency, or equivalently larger detuning, as $L$ increases.

\subsubsection{$\mathbb{D}_3$ symmetry breaking}

The transitions from the disordered phase to the $\mathbb{D}_3$ phase and from the $\mathbb{Z}_2^+$ phase to the $\mathbb{Z}_2\times \mathbb{D}_3$ phase are both associated with spontaneous breaking of the $\mathbb{D}_3$ leg-exchange symmetry. To characterize this symmetry breaking, we introduce a complex local field
\begin{equation}
\eta_i=\frac{1}{3}\sum_{s=1}^{3} e^{\mathrm{i}\frac{2\pi}{3}(s-1)} \langle \hat{n}_{i,s}\rangle,
\end{equation}
which measures the density imbalance among the three legs of the $i$th triangle. The corresponding order parameter is defined by its Fourier component at the appropriate ordering wave vector,
\begin{equation}
\Psi_q=\frac{1}{L}\left|\sum_j e^{\mathrm{i}qj}\eta_j\right|.
\end{equation}
For the $\mathbb{D}_3$ and $\mathbb{Z}_2\times\mathbb{D}_3$ phases discussed here, the relevant wave vectors are $q=2\pi/3$ and $q=\pi/3$, respectively.

\begin{figure}[t]
    \centering
    \includegraphics[width=1\linewidth]{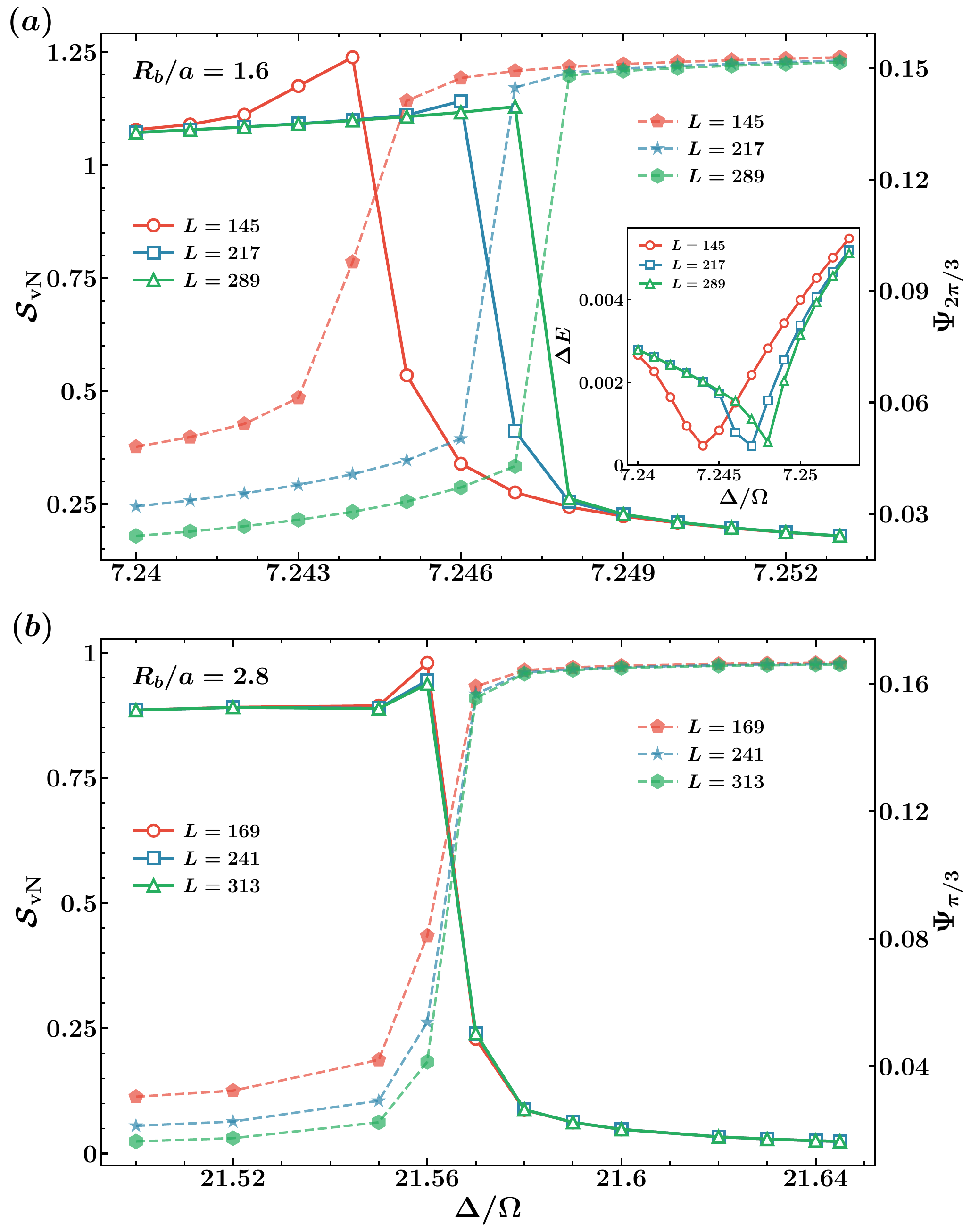}
    \caption{The von Neumann entanglement entropy $\mathcal{S}_{\rm vN}$ and the order parameter as functions of $\Delta/\Omega$ across phase transitions associated with spontaneous $\mathbb{D}_3$ symmetry breaking. (a) Results along the cut $R_b/a=1.6$, crossing from the disordered phase to the $\mathbb{D}_3$ phase. The open markers denote $\mathcal{S}_{\rm vN}$ and correspond to the left axis, while the filled markers denote the order parameter $\Phi_{2\pi/3}$ and correspond to the right axis. The inset shows that the energy gap as a function of $\Delta/\Omega$ for different system sizes. (b) Same as in (a), but along the cut $R_b/a=2.8$, crossing from the $\mathbb{Z}_2^+$ phase to the $\mathbb{Z}_2\times\mathbb{D}_3$ phase. Here the filled markers denote the order parameter $\Phi_{\pi/3}$. %The finite discontinuities in $\mathcal{S}_{\rm vN}$, $\Phi_{2\pi/3}$, and $\Phi_{\pi/3}$, together with the behavior of the energy gap, indicate that the transitions associated with $\mathbb{D}_3$ symmetry breaking are first order.
    }
    \label{fig:SvN_Order_Rba}
\end{figure}

The numerical results are shown in Fig.~\ref{fig:SvN_Order_Rba}. Along the cut $R_b/a=1.6$, which crosses from the disordered phase to the $\mathbb{D}_3$ phase, both $\mathcal{S}_{\rm vN}$ and the order parameter $\Psi_{2\pi/3}$ exhibit clear jumps as functions of $\Delta/\Omega$. Along the cut $R_b/a=2.8$, which crosses from the $\mathbb{Z}_2^+$ phase to the $\mathbb{Z}_2\times\mathbb{D}_3$ phase, the same behavior is observed for $\mathcal{S}_{\rm vN}$ and the corresponding order parameter $\Psi_{\pi/3}$. In both cases, the jump in the order parameter does not decrease with increasing system size. At the same time, $\mathcal{S}_{\rm vN}$ does not develop a size-growing peak or any divergent regime near the transition. For the $R_b/a=1.6$ cut in Fig.~\ref{fig:SvN_Order_Rba}(a), we also show the energy gap in the inset. As the system size increases, the gap develops an increasingly sharp minimum near the transition. However, the minimum does not exhibit the $L^{-z}$ scaling expected for a continuous transition. Instead, once it becomes smaller than $10^{-3}$, it no longer decreases systematically with $L$ within the present numerical resolution. This behavior is consistent with a rounded level crossing, where the true minimum can be limited by the finite step size in $\Delta/\Omega$, finite bond dimension, and residual finite-size effects. Taken together, these observations rule out a continuous transition and instead indicate a first-order one.

\begin{figure}[t]
    \centering
    \includegraphics[width=1\linewidth]{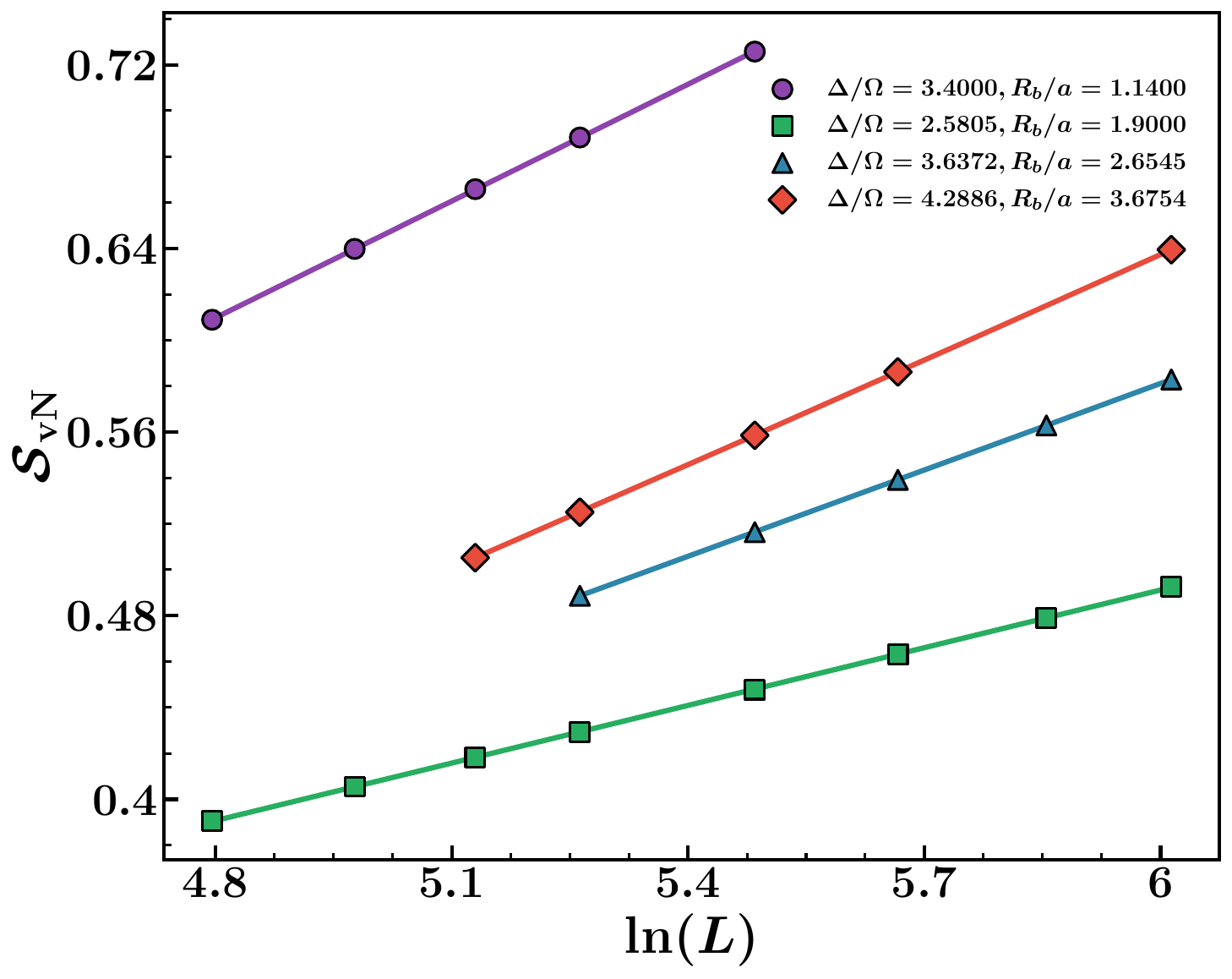}
    \caption{The von Neumann entanglement entropy $\mathcal{S}_{\rm vN}$ as a function of $\ln L$ at the quantum critical point $(\Delta/\Omega,R_b/a)=(3.4,1.14)$ inside the six-state-clock critical phase, the Ising CFT point $(2.5805,1.9)$ on the boundary of the $\mathbb{Z}_2^+$ phase, the Potts CFT point $(3.6372,2.6545)$ on the boundary of the $\mathbb{Z}_3^+$ phase, and the AT CFT point $(4.2886,3.6754)$ on the boundary of the $\mathbb{Z}_4^+$ phase. For clarity, the values of $\mathcal{S}_{\rm vN}$ at $(3.4,1.14)$ are shifted downward by $1$. The data for the first two critical points are fit to the CFT scaling form in Eq.~\eqref{eq:cfteeform}. The extracted central charges are $c\approx 1.023$ and $c\approx 0.505$, respectively. %, in good agreement with the CFT predictions $c=1$ and $c=0.5$. 
    The latter two critical points are fit to Eq.~\eqref{eq:cfteeform} supplemented by an additional correction term proportional to $1/\ln L$. The resulting central charges are $c\approx 0.793$ for the Potts CFT point and $c\approx 1.008$ for the AT CFT point. %, again in good agreement with the CFT predictions $c=0.8$ and $c=1$. 
    The truncation error is set to $\varepsilon=10^{-12}$.}%power-law correction term $a/L^p$ to account for finite-size effects. The fits give $c\approx 0.763$ with $p\approx 1.730$ for the third critical point, and $c\approx 0.957$ with $p\approx 0.749$ for the fourth critical point.}
    \label{fig:CenChar}
\end{figure}

\begin{figure}[t]
    \centering
    \includegraphics[width=1\linewidth]{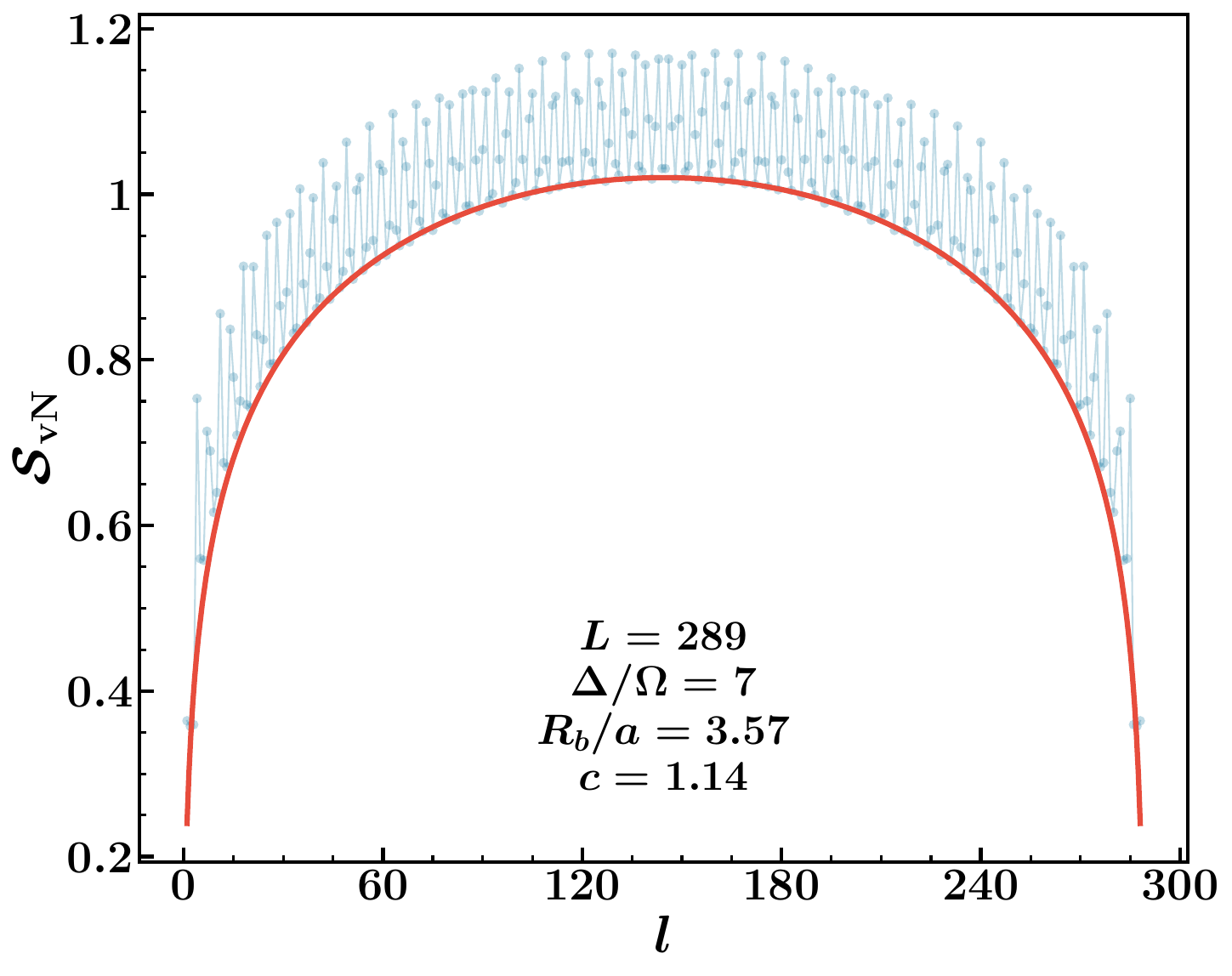}
    \caption{The entanglement entropy $\mathcal{S}_{\rm vN}$ as a function of subsystem size $l$ in the gapless floating phase located between the $\mathbb{Z}_3^+$ and $\mathbb{Z}_4^+$ ordered phases at $\Delta/\Omega=7$ and $R_b/a=3.57$. The middle $2/3$ of the data points on the lower envelope are fit to Eq.~\eqref{eq:cfteeform} to extract the central charge. The truncation error is set to $\varepsilon=10^{-12}$.}
    \label{fig:FloatFit}
\end{figure}

\subsubsection{Central charges at critical points and gapless phases}

To further confirm the universality classes of the continuous phase transitions in the triangular-prism Rydberg array, we analyze the scaling of the von Neumann entanglement entropy $\mathcal{S}_{\rm vN}$ and extract the central charge by fitting the center-cut entropy to Eq.~\eqref{eq:cfteeform}. The results for representative critical points on the phase boundaries are summarized in Fig.~\ref{fig:CenChar}. For the transition between the disordered phase and the $\mathbb{Z}_2^+$ phase at $(\Delta/\Omega,R_b/a)\approx(2.5805,1.9)$, the fit gives $c\approx 0.505$, in excellent agreement with the Ising CFT prediction $c=1/2$. For the direct melting of the $\mathbb{Z}_3^+$ and $\mathbb{Z}_4^+$ phases, a direct fit to Eq.~\eqref{eq:cfteeform} shows a noticeable finite-size drift. To obtain more reliable estimates, we include an additional correction term proportional to $1/\ln L$. Such logarithmic corrections are natural near the four-state Potts limit of the Ashkin-Teller theory \cite{CardyLogarithmic1986,LiuBoundaryCritical2026}. At the Potts CFT point $(\Delta/\Omega,R_b/a)=(3.6372,2.6545)$, this gives $c\approx 0.793$, consistent with the three-state Potts value $c=0.8$. At the $\mathbb{Z}_4^+$ CFT point $(\Delta/\Omega,R_b/a)=(4.2886,3.6754)$, the same fitting form gives $c\approx 1.008$, in excellent agreement with the AT CFT value $c=1$. For comparison, fitting without corrections gives $c\approx 0.753$ and $0.911$ for the Potts and AT points, respectively, while fitting with a power-law correction term $1/L^b$ gives $c\approx 0.762$ and $0.956$. These values are also consistent with the expected CFT predictions, but the $1/\ln L$ correction gives the closest agreement in our data.

For the intermediate critical phase associated with the melting of the $\mathbb{Z}_2\times\mathbb{Z}_3$ order, we also show the result at $(\Delta/\Omega,R_b/a)=(3.4,1.14)$ in Fig.~\ref{fig:CenChar}. Fitting with Eq.~\eqref{eq:cfteeform} without additional corrections gives $c\approx 1.023$, again consistent with the expectation $c=1$ for the Gaussian critical theory underlying the $\mathbb{Z}_6$ clock physics. We also analyze the entanglement entropy in the floating phase between the $\mathbb{Z}_3^+$ and $\mathbb{Z}_4^+$ phases. In Fig.~\ref{fig:FloatFit}, we plot $\mathcal{S}_{\rm vN}$ as a function of subsystem size $l$ at fixed $L=289$ and $(\Delta/\Omega,R_b/a)=(7,3.57)$. The entropy shows pronounced oscillations with a large envelope due to the incommensurability, similar to those found in ladder systems \cite{ZhangFloating2025NC,LiaoPhaseRydLadder2025prb}. Fitting the center $2/3$ data points on the lower envelope to Eq.~\eqref{eq:cfteeform} gives $c\approx 1.14$, which is also consistent with the Gaussian CFT expectation $c=1$. Taken together with the Binder-cumulant and gap-scaling analyses discussed above, these central-charge estimates provide a consistent characterization of the critical structure of the triangular-prism Rydberg array.

\begin{figure}[b]
    \centering
    \includegraphics[width=1\linewidth]{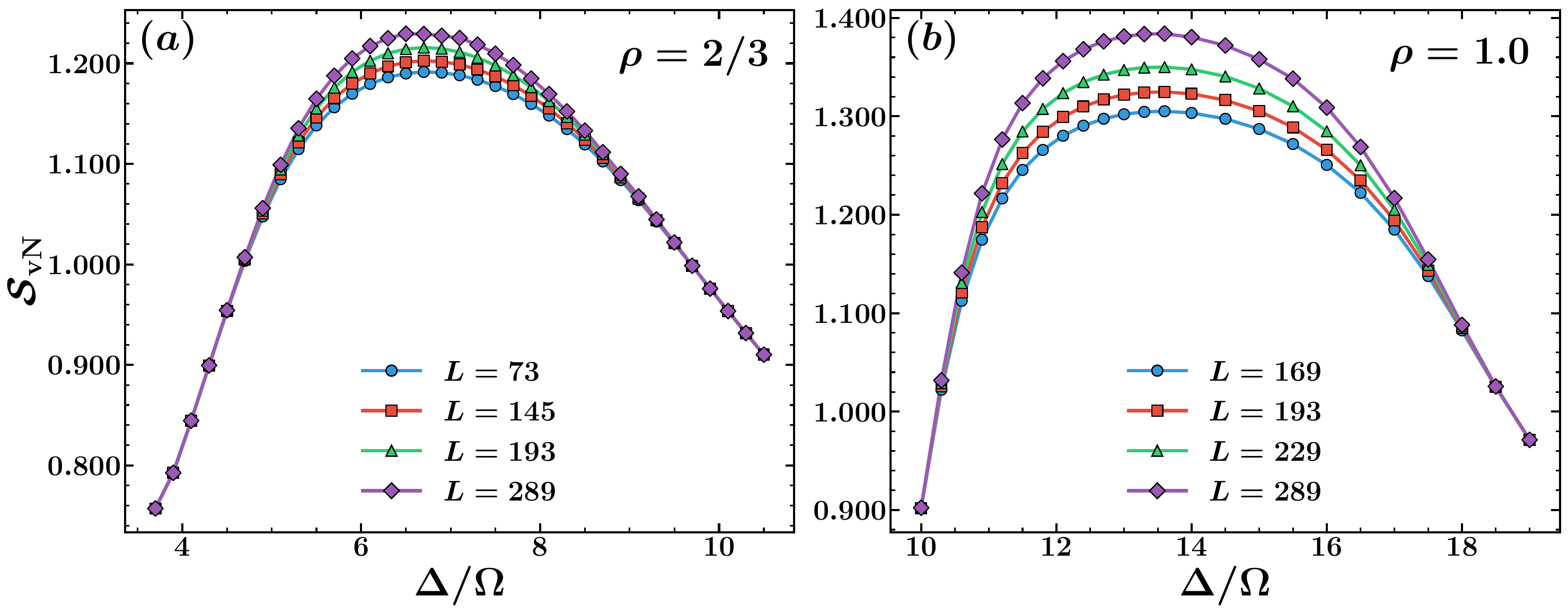}
    \caption{The entanglement entropy $\mathcal{S}_{\rm vN}$ as a function of $\Delta/\Omega$ for different system sizes in the high-entanglement regime inside the $\mathbb{Z}_2^+$ phase. (a) Results for aspect ratio $\rho=a_x/a=2/3$, taken along the cut $R_b/a=2.4+(\Delta/\Omega-3.6)/10$. (b) Results for $\rho=a_x/a=1$, taken along the cut $R_b/a=1.94+(\Delta/\Omega-10)/10$. The truncation error is set to $\varepsilon=10^{-11}$.}
    \label{fig:SvNdiffrho}
\end{figure}

\begin{figure}[t]
    \centering
    \includegraphics[width=1\linewidth]{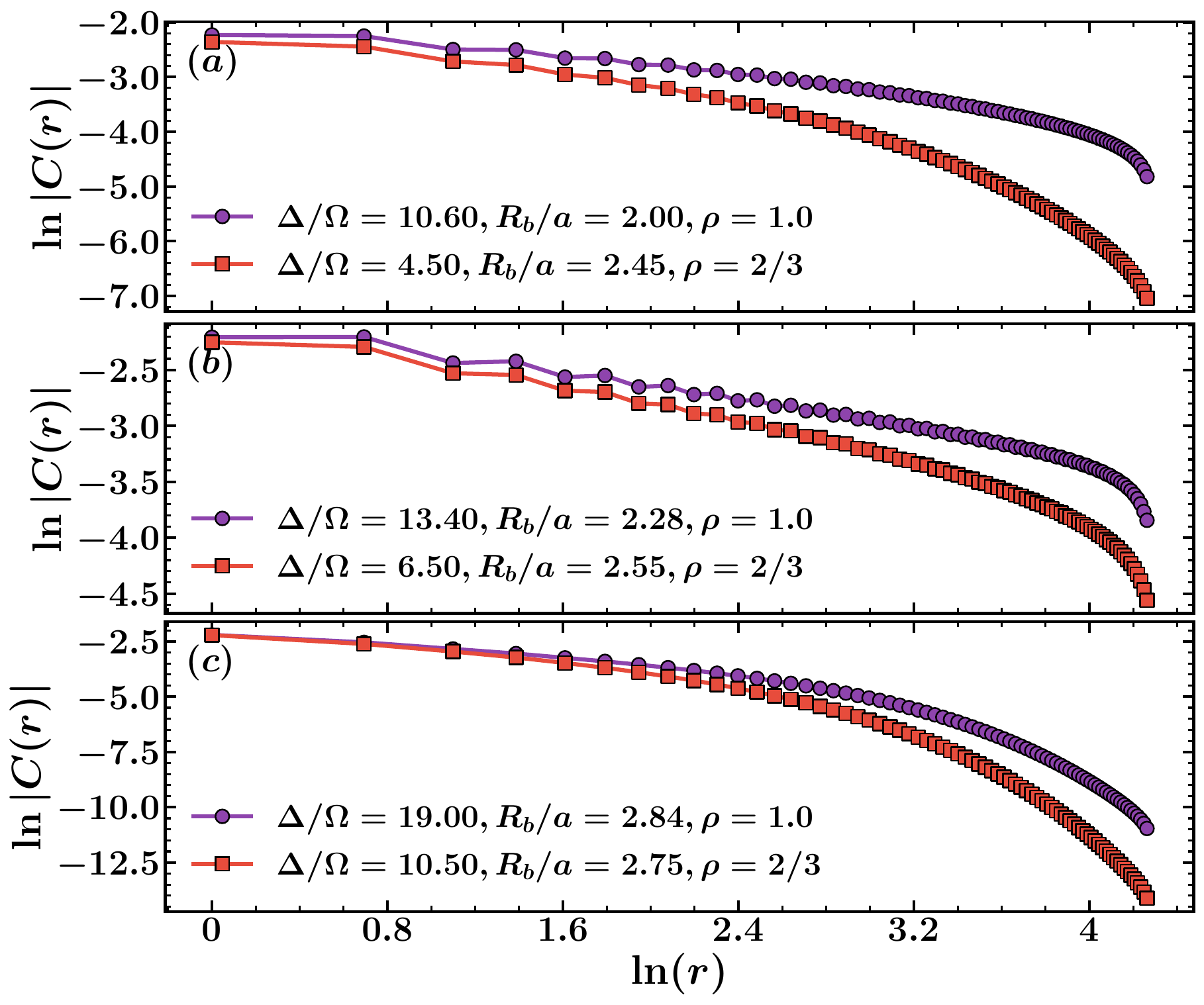}
    \caption{Log-log plots of the connected density-density correlation function $|C(r)|$ as a function of distance $r$ along the first leg, restricted to the occupied-triangle sublattice inside the $\mathbb{Z}_2^+$ phase. The three panels correspond to different regimes relative to the entanglement-entropy peak shown in Fig.~\ref{fig:SvNdiffrho}: (a) to the left of the peak, (b) at the peak position, and (c) to the right of the peak. Each panel compares two geometric configurations, $\rho=1$ and $\rho=2/3$.}
    \label{fig:Corrdifrho}
\end{figure}

\subsection{Crossover regime inside the $\mathbb{Z}_2^+$ phase}
\label{subsec:crossover}
As discussed in Sec.~\ref{subsec:phasediagram}, the $\mathbb{Z}_2^+$ lobe contains a high-entanglement regime before the system eventually enters the $\mathbb{Z}_2\times \mathbb{D}_3$ phase at larger detuning. The question is whether this feature should be viewed as a crossover or as the precursor of a genuine phase transition. Physically, this regime originates from the increasing repulsion between occupied triangles inside the $\mathbb{Z}_2^+$ phase. As $\Delta/\Omega$ increases, the system develops a tendency toward an additional period-2 modulation within the occupied-triangle sublattice. In the prism geometry, however, this tendency competes with the eventual $\mathbb{D}_3$ symmetry breaking at larger detuning, which instead produces an additional period-3 modulation on each leg. To examine how this competition depends on geometry, we compare the von Neumann entanglement entropy and the connected density-density correlations for two aspect ratios, $\rho=a_x/a=1$ and $2/3$.

Figure~\ref{fig:SvNdiffrho} shows $\mathcal{S}_{\rm vN}$ as a function of $\Delta/\Omega$ for different system sizes inside the $\mathbb{Z}_2^+$ phase. For both aspect ratios, an entropy peak appears inside the lobe. Its size dependence, however, is qualitatively different in the two cases. For $\rho=2/3$, the peak saturates quickly with increasing $L$, which is more consistent with a crossover. By contrast, for $\rho=1$, the peak grows more visibly with system size, indicating that the system is closer to criticality. Nevertheless, on both sides of the peak, $\mathcal{S}_{\rm vN}$ decreases smoothly and becomes nearly size independent, which indicates that these neighboring regimes remain gapped. The peak region is also smoothly connected to them, with no sign of a discontinuity that would indicate an additional thermodynamic phase transition. In addition, throughout this regime the occupied triangles still have local density close to $1/3$, so there is no evidence for a new symmetry-breaking phase emerging inside the $\mathbb{Z}_2^+$ lobe.

To further characterize this regime, we compute the connected density-density correlation function
\begin{equation}
C(r)=\langle \hat{n}_{i,s}\hat{n}_{i+r,s}\rangle-\langle \hat{n}_{i,s}\rangle\langle \hat{n}_{i+r,s}\rangle
\end{equation}
along the $s=1$ leg. The results for the occupied triangles are shown in Fig.~\ref{fig:Corrdifrho}. On both sides of the entropy peak, the correlations decay rapidly, again consistent with gapped behavior. Near the peak position, however, the correlations decay much more slowly and develop a clear additional period-2 modulation within the occupied subsystem. For $\rho=1$, the decay near the peak is much closer to an algebraic form over the accessible distances, whereas for $\rho=2/3$ it still bends downward more clearly at long distance. This is consistent with the larger $\mathcal{S}_{\rm vN}$ found for $\rho=1$. Within the parameter range studied here, we therefore do not identify a separate ordered phase inside the $\mathbb{Z}_2^+$ lobe. The high-entanglement regime is best viewed as a crossover, although its strong aspect-ratio dependence indicates that it may become substantially sharper as the geometry is varied.

%%%%%%%%%%%%%%%%%%%%%%%%%%%%%%%%%%%%%%%%%%%%%%%%%%%%%%%%%%%%%%%%%%%%%%%%%%
\section{Conclusions}\label{sec:conclusion}
%%%%%%%%%%%%%%%%%%%%%%%%%%%%%%%%%%%%%%%%%%%%%%%%%%%%%%%%%%%%%%%%%%%%%%%%%%

In this work, we systematically mapped the ground-state phase diagram of a Rydberg atom array in a triangular-prism geometry using the density matrix renormalization group method. The interplay among the blockade constraint, translation symmetry along the prism axis, and the $\mathbb{D}_3$ leg-exchange symmetry of the triangular cross section gives rise to a rich set of crystalline phases and quantum phase transitions. Besides the trimerized $\mathbb{Z}_p^+$ phases, we identified phases with partial or complete $\mathbb{D}_3$ symmetry breaking, including the $\mathbb{Z}_2\times\mathbb{Z}_3$, $\mathbb{D}_3$, and $\mathbb{Z}_p\times\mathbb{D}_3$ orders.

For the $\mathbb{Z}_p^+$ phases, the physics is largely the same as that in Rydberg chains and ladders \cite{MaceiraChiral2022prr,eck2023critical,ZhangFloating2025NC,LiaoPhaseRydLadder2025prb,ChepigaLadder2025prr}. In particular, the transition from the disordered phase to the $\mathbb{Z}_2^+$ phase belongs to the Ising universality class, while the direct melting of the $\mathbb{Z}_3^+$ and $\mathbb{Z}_4^+$ phases is controlled by isolated Potts and Ashkin-Teller CFT points on the phase boundaries, with chiral transitions away from those points and floating phases beyond the Lifshitz points. At the same time, the prism geometry brings additional multi-leg effects that are absent in chains and ladders. In the relatively small-$R_b/a$ regime, where experiments are easier to access, we find the $\mathbb{Z}_2\times\mathbb{Z}_3$ phase with an intermediate critical phase described by effective $\mathbb{Z}_6$ clock physics and bounded by two BKT transitions. In the same regime, we also find the $\mathbb{D}_3$ phase and show that the transitions associated with $\mathbb{D}_3$ symmetry breaking are first order. In this sense, the multi-leg geometry moves nontrivial physics into a parameter regime that should be more accessible experimentally, similar in spirit to how the two-leg geometry makes the floating phase easier to explore \cite{ZhangFloating2025NC}.

We also found that the $\mathbb{Z}_2^+$ lobe contains a high-entanglement regime before the system enters the $\mathbb{Z}_2\times\mathbb{D}_3$ phase at larger detuning. Our analysis of the entanglement entropy and connected density-density correlations shows that, within the parameter range studied here, this regime is best understood as a crossover rather than a separate phase. Physically, it reflects the tendency toward an additional period-2 modulation inside the occupied-triangle sublattice, which competes with the eventual $\mathbb{D}_3$ symmetry breaking at larger detuning. Such a crossover should in principle appear inside every $\mathbb{Z}_p^+$ lobe, although within the present parameter range it is most clearly visible in the $\mathbb{Z}_2^+$ phase.

Taken together, these results establish the triangular-prism Rydberg array as a versatile setting for studying how nontrivial cross-sectional symmetry enriches the phase structure and criticality of quasi-one-dimensional Rydberg systems. Our work provides concrete predictions for future experiments in programmable tweezer arrays, especially in the small-$R_b/a$ regime where the $\mathbb{Z}_6$ clock critical phase and the first-order $\mathbb{D}_3$ transitions should be accessible. An important direction for future work is to clarify the full effect of varying the aspect ratio $a_x/a$. In particular, it remains open whether the crossover regimes inside the $\mathbb{Z}_p^+$ lobes can sharpen into genuine criticality for suitable geometry.

\begin{acknowledgments}
We thank Li-Ping Yang and Shu-Ao Liao for helpful discussions. J.Z. acknowledges support from the National Natural Science Foundation of China under Grant No. 12304172 and from the Chongqing Natural Science Foundation under Grant No. CSTB2024YCJH-KYXM0064. This work was also supported in part by the National Natural Science Foundation of China under Grant No. 12547101. Part of the computations were performed using the computer clusters and data storage resources of the HPCC, which were funded by grants from NSF (MRI-2215705, MRI-1429826) and NIH (1S10OD016290-01A1).
\end{acknowledgments}

\end{document}